%% file: papers81.tex
\documentclass[12pt] {article}
\usepackage{latexsym}{\rm }
\usepackage{graphics}
\usepackage{graphpap}
\input{Definiti}
\input{Leaf1}
\begin{document}

%%\color{black}
\begin{center}

%\vspace{2cm}

{\LARGE \bf The $d=6$ trace anomaly from quantum field theory four-loop graphs in one dimension.}\\ [5mm]

\large{Agapitos Hatzinikitas}\footnote [2] 
{Email: ahatzini@tem.uoc.gr} \\ [5mm]

{\small University of Crete, \\
Department of Applied Mathematics, \\
L. Knosou-Ambelokipi, 71409 Iraklio Crete,\\ 
Greece \\
and \\
University of Athens, \\
Nuclear and Particle Physics Division,\\
Panepistimioupoli GR-15771 Athens, Greece}\\ [5mm]

\large{and Renato Portugal} \footnote [3] 
{Email: portugal@Lncc.br}\\ [5mm]

{\small Laborat\'orio Nacional de
Computa\c{c}\~ao Cient\'{\i}fica,\\ 
Av. Getulio Vargas, 333,\\ 
Petr\'opolis, RJ, Brazil. Cep 25651-070.}
\vspace{5mm}

\end{center}

\begin{abstract}

We calculate the integrated trace anomaly for a real spin-0 scalar field in six dimensions
in a torsionless curved space without a boundary. We use a  path integral approach for a corresponding supersymmetric 
quantum mechanical model.
Weyl ordering the corresponding 
Hamiltonian in phase space, an extra two-loop counterterm 
$\frac{1}{8}\bigg(R + g^{ij} \Gamma^{l}_{k i} \Gamma^{k}_{l j} \bigg)$ is produced in the action.
Applying a recursive method we evaluate the components of the metric tensor in Riemann normal coordinates in  
six dimensions and construct the interaction Langrangian density by employing the background field method. 
 The calculation 
of the anomaly is based on the end-point scalar propagator and not on the 
string inspired center-of-mass propagator which gives incorrect 
results for the local trace anomaly.  
The manipulation of the Feynman diagrams is partly relied on the factorization of four dimensional subdiagrams 
and partly on a brute force computer 
algebra program developed to serve this specific purpose. The computer program enables one to perform index contractions of
twelve quantum fields (10395 in the present case) a task which cannot be accomplished otherwise.
We observe that the contribution of the disconnected diagrams is no longer 
proportional to the two dimensional trace anomaly (which vanishes in four dimensions). 
The integrated trace anomaly is finally expressed in terms of the 17 linearly independent scalar 
monomials constructed out of covariant derivatives and Riemann tensors.

\end{abstract}



%%%%%%%%%%%%%%%%%%%%%%%%%%%%%%%%%%%%%%%%%%%%%%%%%%%%%%%%%%%%%%%%%%%%%%%%%%%%%%%%%%%%%%%%%%%%%%%%%%%%%%%%%%%%%%%%%%%%%%%

						%%%%%%%%%%%%%%%%%%
						%  INTRODUCTION  %
						%%%%%%%%%%%%%%%%%%
\newpage

\section{Introduction}
\renewcommand{\thefootnote}{\arabic{footnote}}
\setcounter{footnote}{0}

The trace anomaly, namely the breaking of classical conformal invariance of gravity actions under Weyl rescaling of 
the metric:
\beqr
g_{ij}(x) \rightarrow \Omega(x) g_{ij}(x),
\label{mere}
\feqr

\noi has a long history with numerous applications and implications to high energy physics, 
general relativity and statistical mechanics. The literature on this subject is vast and in this brief review we will
only concentrate on those aspects which are associated with the present problem. For a historical review the reader
is advised to consult \cite{duff0}. 

Interesting different methods have been developed to investigate and calculate Weyl anomaly in four and 
higher dimensions. The authors in \cite{duff} were able to express the integrated trace anomaly in four dimensions as a 
linear combination of two invariants, the square of Weyl tensor: 
\beqr
C_{ijkl}C^{ijkl}= R_{ijkl}^{2}-2R_{ij}^{2}+ \frac{1}{3}R^{2}
\label{quweyl}
\feqr

\noi and the only parity even candidate:  
\beqr
\mathcal{E}_4 ={}^{\ast}R_{ijkl} {}^{\ast}R^{ijkl}= 
R_{ijkl}^{2}-4R_{ij}^{2}+R^{2}
\label{eu4}
\feqr

\noi which is proportional to the well-known Gauss-Bonnet topological density and ${}^{\ast}$ denotes the dual. 
To be more concrete, the gravitational contribution to the anomaly depends on only two constants (call them $\alpha$ and 
$\beta$) and is expressed as
\beqr
g^{ij}<T_{ij}>=\alpha \left(C^{ijkl} C_{ijkl}+ \frac{2}{3} \Box R \right)
+ \beta \mathcal{E}_4.
\label{grav4}
\feqr

\noi The numerical values of the constants are: $\alpha=\frac{1}{\pi^2} \frac{1}{30}\frac{1}{64}$, 
$\beta=-\frac{1}{\pi^2} \frac{1}{90}\frac{1}{64}$ and can also be found using the Feynman diagram scheme in \cite{takis}.

It was soon realised \cite{dewitt} 
that these invariants were manifested in the t-independent 
$b_{2}$ coefficient of the Schwinger-De Witt asymptotic expansion of the heat kernel of the appropriate 
differential operators.
The four (and partially the six) dimensional anomaly was later rederived by Bonora et. al. \cite{bonora} 
who established the connection between Weyl anomalies and cocycles by relying on a cohomological method 
and using the Wess-Zumino consistency condition. Although these authors explicitly specified the invariants
(see Appendix A.2 for their classification) they did
 not express the six dimensional trace anomaly in terms of these invariants. 
From the representation theory point of view Fulling et. al. \cite{fulling} were also able to determine the 
number of independent scalar monomials of each order and degree up to twelve in derivatives of the metric. 
The first explicit result was given in the literature by Gilkey \cite{gilkey} and later by Avramidi 
\cite{avramidi}, who by 
making some modifications and innovations to the ``heat kernel'' or ``proper time'' method, presented a new covariant 
nonrecursive procedure and found the one-loop effective action in the presence of arbitrary background fields in 
six and eight curved space-time dimensions. Later Deser and Schwimmer \cite{deser} re-examined the diferrent origin of 
the topological (type-A) versus local conformal scalar polynomials involving powers of the Weyl tensor and its 
derivatives 
(type-B) contributions to the anomaly in general dimensions.

In the present work we follow the supersymmetric quantum mechanical approach first pioneered by 
Alvarez-Gaum\'{e} and Witten \cite{alvarez1} and used 
to compute chiral anomalies. According to this method the operators $\gamma_5$,
 $\nabla_{\mu}$, $x_{\mu}$, $\gamma^{\mu}$ 
are represented by operators of a corresponding quantum mechanical model, and by 
turning these operator expressions into path integrals, one finds that anomalies of 
quantum field theories can be written in terms of Feynman diagrams for certain sigma 
models on the worldline. Bastianelli and van Nieuwenhuizen applied this method to trace anomalies 
\cite{peter}. These authors used mode regularization, a scheme widely used at the time. Subsequent 
work by de Boer et. al \cite{kostas} showed how to use time-slicing and gave a completely and unambigous derivation 
of trace anomalies in terms of path integrals, using as input Einstein Hamiltonians. In this article we apply the 
regularization method of \cite{kostas} to the calculation of trace anomalies, following the set up of 
\cite{peter}. 

The outline of the paper is as follows. In the next section we present explicitly one of the basic ingredients of the
background field method, namely the expansion of the metric tensor components in six dimensions in Riemann normal coordinates. 
The method we use 
is recursive and enables one with the application of \rf{genex} and \rf{coeexp} to determine the expansion of the metric 
tensor up to the desired order.

Section 3 begins with a very rapid introduction to trace anomalies from the one-dimensional path integral point of view. We 
write down the interaction Lagrangian density and the propagators of the fields involved. 

Section 4 is devoted to the calculation of the perturbative expansion depicting at the same time the Feynman 
diagrams associated to each vertex.  

A vital tool in our journey is a computer algebra algorithm which proved to be very efficient especially in finding
the contribution of $I_6$ vertex with twelve quantum fields. Thus section 5 is devoted to a brief description of this 
program. We illustrate it's capabilities by applying it to the $I_7$ interaction with eight quantum fields.

Our conclusions are given in section 6. Several appendices follow to assist to a deeper understanding of the 
technical obstructions the reader might face.
%%%%%%%%%%%%%%%%%%%%%%%%%%%%%%%%%%%%%%%%%%%%%%%%%%%%%%%%%%%%%%%%%%%%%%%%%%%%%%%%%%%%%%%%%%%%%%%%%%%%%%%%%%%%%%%%%%%%%%%%

						%%%%%%%%%%%%%%%%
						%   SECTION 1  %
						%%%%%%%%%%%%%%%%

\section{The recursive expansion of the metric components in RNC in six space-time dimensions}

Before embarking on the background field method we discuss the expansion of the metric tensor in Riemann normal coordinates 
(RNC). 

RNC have the appealing feature that the geodesics passing through the origin have the same form as the equations of 
straight 
lines passing through the origin of a Cartesian system of coordinates in Euclidean geometry \cite{petrov}. 
Locally no two geodesics through 
a point $\mathcal{P}$ intersect at another point, and the power series solution of the geodesic equation is:
\beqr
y^l = \xi^{i_1} s + \sum_{k=2}^{\infty} \frac{1}{k!} 
\left( \bar{\Gamma}^{l}_{i_1 i_2 \cdots i_k} \right)_{\mathcal{P}} 
\xi^{i_1} \xi^{i_2} \cdots \xi^{i_k} s^k.
\label{riepo}
\feqr

\noi where $\left(\Gamma^{l}_{i_1 i_2 \cdots i_k}\right)_{\mathcal{P}}$ are the 
$\textit{``generalized Christoffel symbols''}$ 
at the point $\mathcal{P}$ and the geodesics through $\mathcal{P}$ which are straight lines are defined in terms of the arc 
length s by:
\beqr
y^l=\xi^l s.
\label{strai}
\feqr

\noi By induction, one can easily prove that:
\beqr
\partial_{(i_1} \partial_{i_2} \cdots \partial_{i_{k-2}} 
\bar{\Gamma}^{l}_{i_{k-1} i_{k} )} =0
\label{eqre}
\feqr

\noi Paraphrasing eq. \rf{eqre}, one can state that all symmetric derivatives of the affine connection vanish at 
the origin in RNC.
 
In general a covariant second rank tensor field on a manifold can be expanded according to:
\beqr
T_{k_1 k_2}(\tilde{\phi})=T_{k_1 k_2}(\phi)+\sum_{n=1}^{\infty} \frac{1}{n!} \left[
\frac{\partial}{\partial \xi_{i_1} } \frac{\partial}{\partial \xi_{i_2}}
\cdots \frac{\partial}{\partial \xi_{i_n}}\right] T_{k_1 k_2}(\phi)
\xi_{i_1}\xi_{i_2} \cdots \xi_{i_n}.
\label{exp}
\feqr

\noi The coefficients of the Taylor expansion are tensors and can be expressed in terms of the components $R^{l}_{m n p}$
 of the Riemann curvature tensor and the covariant derivatives $D_{k} T_{lm}$ and $D_{k}R^{l}_{m n p}$. Without 
much effort one can prove that:
\beqr
\partial_{(i_1} \partial_{i_2} \cdots \partial_{i_{n-1}} 
\bar{\Gamma}^{l}_{i_n ) k} = &-& \left(\frac{n-1}{n+1} \right) \Big[
D_{(i_1} D_{i_2} \cdots D_{i_{n-2}} \bar{R}^{l}_{i_{n-1}k i_n)} \non \\
&+& \partial_{(i_1} \partial_{i_2} \cdots \partial_{i_{n-2}} \left
(\bar{\Gamma}^{\alpha}_{i_{n-1}k} \bar{\Gamma}^{l}_{\alpha i_n )} 
\right) \non \\
&-& \partial_{(i_1} \partial_{i_2} \cdots \partial_{i_{n-3}}
\left(\bar{\Gamma}^{l}_{i_{n-2} \alpha} \bar{R}^{\alpha}_{i_{n-1}ki_n )} 
- l \leftrightarrow \alpha, \alpha \leftrightarrow k \right) \non \\
&-& \partial_{(i_1} \partial_{i_2} \cdots \partial_{i_{n-4}}
\left(\bar{\Gamma}^{l}_{i_{n-3} \alpha} D_{i_{n-2}}
\bar{R}^{\alpha}_{i_{n-1}ki_n )} 
- l \leftrightarrow \alpha, \alpha \leftrightarrow k \right) \non \\
&\vdots& \non \\
&-& \left( \partial_{(i_1} \bar{\Gamma}^{l}_{i_2 \alpha} D_{i_3} \cdots D_{i_{n-2}}
\bar{R}^{\alpha}_{i_{n-1} k i_n )} - l \leftrightarrow \alpha, \alpha \leftrightarrow k
\right) \Big]
\label{genex}
\feqr

\noi where the interchange of covariant and contravariant indices act independently and symmetrization
acts only on i indices. Expression \rf{genex} reproduces 
for various values of n\footnote{In \cite{alvarez} there is a misprint for the $n=4$ case. A minus sign is needed in 
front of the $\frac{2}{9}$-term.} the following results:
\beqr
\partial_{(i_1}\bar{\Gamma}^{l}_{i_2 )  k} = &-&  \frac{1}{3} 
\bar{R}^{l}_{(i_1 k i_2)} \non \\
\partial_{(i_1}\partial_{i_2}\bar{\Gamma}^{l}_{i_3) k} = 
&-&\frac{1}{2}D_{(i_1}\bar{R}^{l}_{i_2 k i_3 )} \non \\
\partial_{(i_1}\partial_{i_2} \partial_{i_3}
\bar{\Gamma}^{l}_{i_4) k} = 
&-&\frac{3}{5} \left[D_{(i_1} D_{i_2} \bar{R}^{l}_{i_3 k i_4 )}
+ \frac{2}{9} \bar{R}^{l}_{(i_1 i_2 \alpha} \bar{R}^{\alpha}_{i_3 i_4 )k}
\right] \non \\
\partial_{( i_1} \partial_{i_2} \partial_{i_3} \partial_{i_4}
\bar{\Gamma}^{l}_{i_5) k} =
&-&\frac{2}{3} \left[D_{(i_1} D_{i_2} D_{i_3} \bar{R}^{l}_{i_4 k i_5 )}
- D_{(i_1}\bar{R}^{\alpha}_{i_2 k i_3} \bar{R}^{l}_{i_4 i_5 )\alpha} 
\right]\non \\
\partial_{( i_1} \partial_{i_2} \partial_{i_3} \partial_{i_4}
\partial_{i_5} \bar{\Gamma}^{l}_{i_6) k} =  
&-&\frac{5}{7} \Big[D_{(i_1} \cdots D_{i_4} \bar{R}^{l}_{i_5 k i_6 )} \non \\
&-& \frac{1}{5} \left(7D_{(i_1}D_{i_2}\bar{R}^{l}_{i_3 \alpha i_4} 
\bar{R}^{\alpha}_{i_5 i_6 )k } + D_{(i_1}D_{i_2}\bar{R}^{\alpha}_{i_3 k i_4} 
\bar{R}^{l}_{i_5 i_6 ) \alpha } \right) \non \\
&+& \frac{3}{2}D_{(i_1} \bar{R}^{\alpha}_{i_2 k i_3}
 D_{i_4} \bar{R}^{l}_{i_5 \alpha i_6 )} - \frac{16}{45}
\bar{R}^{l}_{(i_1 i_2 \alpha} \bar{R}^{\alpha}_{i_3 i_4 \beta}
\bar{R}^{\beta}_{i_5 i_6 )k}\Big] \non \\
\cdots
\label{resgen}
\feqr 

\noi The coefficients of\rf{exp} can be rewritten as:
\beqr
\partial_{( i_1} \partial_{i_2} \cdots \partial_{i_n )} \bar{T}_{k_1 k_2} =
& & D_{(i_1}D_{i_2} \cdots D_{i_n )}\bar{T}_{k_1 k_2} \non \\ 
&+& \partial_{( i_1} \partial_{i_2} \cdots \partial_{i_{n-1} )}
\left[\bar{\Gamma}^{\alpha}_{i_n k_1 )} \bar{T}_{\alpha k_2} + 
k_1 \leftrightarrow k_2 \right] \non \\
&+& \partial_{( i_1} \partial_{i_2} \cdots \partial_{i_{n-2} )}
\left[\bar{\Gamma}^{\alpha}_{i_{n-1} k_1} D_{i_n )}\bar{T}_{\alpha k_2} + 
k_1 \leftrightarrow k_2 \right] \non \\
&\vdots& \non \\
&+& \partial_{( i_1} \partial_{i_2}
\left[\bar{\Gamma}^{\alpha}_{i_3 k_1} D_{i_4} \cdots D_{i_n )}
\bar{T}_{\alpha k_2} + 
k_1 \leftrightarrow k_2 \right] \non \\
&+& \partial_{( i_1}
\left[\bar{\Gamma}^{\alpha}_{i_2 k_1} D_{i_3} \cdots D_{i_n )}
\bar{T}_{\alpha k_2} + k_1 \leftrightarrow k_2 \right]. 
\label{coeexp}
\feqr

\noi Expressions \rf{genex} and \rf{coeexp} compose the building blocks of the current recursive method which produces 
the following results for different values of n:
\beqr
\partial_{( i_1} \partial_{i_2 )} \bar{T}_{k_1 k_2} &=& 
D_{(i_1} D_{i_2)} \bar{T}_{k_1 k_2} - \frac{1}{3} \left[
\bar{R}^{\rho}_{(i_1 k_1 i_2 )} \bar{T}_{\rho k_2} + 
k_1 \leftrightarrow k_2 \right] \non \\
\partial_{( i_1} \partial_{i_2} \partial_{i_3 )} \bar{T}_{k_1 k_2} = & & 
D_{(i_1} D_{i_2} D_{i_3 )}\bar{T}_{k_1 k_2} \non \\
&+& \left[\partial_{(i_1} \partial_{i_2} \bar{\Gamma}^{\rho}_{i_3 ) k_1} 
\bar{T}_{\rho k_2} + 2 \partial_{(i_1} \bar{\Gamma}^{\rho}_{i_2 k_1}
D_{i_3 )} \bar{T}_{\rho k_2} + k_1 \leftrightarrow k_2\right] \non \\
&-& \frac{1}{3} \left(\bar{R}^{\rho}_{(i_1 k_1 i_2} D_{i_3 )} 
\bar{T}_{\rho k_2} + k_1 \leftrightarrow k_2 \right) \non \\
\partial_{( i_1} \partial_{i_2} \partial_{i_3} \partial{i_4 )} \bar{T}_{k_1 k_2} = & & 
D_{(i_1} D_{i_2} D_{i_3} D_{i_4 )} \bar{T}_{k_1 k_2} \non \\
&+& \Big[\partial_{( i_1} \partial_{i_2} \partial_{i_3} \bar{\Gamma}^{\rho}_{i_4 ) k_1} \bar{T}_{\rho k_2}
+ 3 \partial_{( i_1} \partial_{i_2} \bar{\Gamma}^{\rho}_{i_3  k_1}D_{i_4 )} \bar{T}_{\rho k_2} \non \\
&+& 3\partial_{(i_1} \bar{\Gamma}^{\rho}_{i_2 k_1} \left(D_{i_3}D_{i_4 )} \bar{T}_{\rho k_2} - \frac{1}{3} 
\left(\bar{R}^{\sigma}_{i_3 \rho i_4 )} \bar{T}_{\sigma k_2} + \rho \leftrightarrow k_2 \right)\right) +
k_1 \leftrightarrow k_2 \Big] \non \\
\partial_{( i_1} \partial_{i_2} \partial_{i_3} \partial{i_4} \partial_{i_5 )} \bar{T}_{k_1 k_2} = & & 
D_{(i_1}D_{i_2}D_{i_3}D_{i_4}D_{i_5 )} \bar{T}_{k_1 k_2} \non \\
&-& \Big[ \frac{10}{3} \bar{R}^{\alpha}_{(i_1 k_1 i_2} D_{i_3} \cdots D_{i_5 )} \bar{T}_{\alpha k_2} +
5 D_{(i_1}\bar{R}^{\alpha}_{i_2 k_1 i_3}D_{i_4}D_{i_5)}\bar{T}_{\alpha k_2} \non \\
&+& 3D_{(i_1}D_{i_2}\bar{R}^{\alpha}_{i_3 k_1 i_4} D_{i_5)}\bar{T}_{\alpha k_2} + 
\frac{2}{3} D_{(i_1}D_{i_2}D_{i_3} \bar{R}^{\alpha}_{i_4 k_1 i_5)} \bar{T}_{\alpha k_2} \non \\
&-& \frac{2}{3}D_{(i_1} \bar{R}^{\rho}_{i_2 k_1 i_3} \bar{R}^{\alpha}_{i_4 i_5 )\rho} \bar{T}_{\alpha k_2}
+D_{(i_1}\bar{R}^{\alpha}_{i_2 k_1 i_3} \left(\bar{R}^{\rho}_{i_4 \alpha i_5)} \bar{T}_{\rho k_2} + 
\alpha \leftrightarrow k_2 \right) \non \\
&+& \frac{2}{3} \left(D_{(i_1}\bar{R}^{\rho}_{i_2 \alpha i_3} \bar{T}_{\rho k_2}+ \alpha \leftrightarrow k_2 \right)
\bar{R}^{\alpha}_{i_4 i_5) k_1} \pm k_1 \leftrightarrow k_2\Big] \non \\
\partial_{( i_1} \partial_{i_2} \partial_{i_3} \partial{i_4} \partial_{i_5} \partial_{i_6)}\bar{T}_{k_1 k_2} = & & 
D_{(i_1} \cdots D_{i_6)} \bar{T}_{k_1 k_2} \non \\
&+& \Big[ \partial_{( i_1} \cdots \partial_{i_5} \bar{\Gamma}^{\alpha}_{i_6) k_1} \bar{T}_{\alpha k_2} 
+ 6  \partial_{( i_1} \cdots \partial_{i_4} \bar{\Gamma}^{\alpha}_{i_5 k_1} D_{i_6)}\bar{T}_{\alpha k_2} \non \\
&+& 15\partial_{( i_1} \cdots \partial_{i_3} \bar{\Gamma}^{\alpha}_{i_4 k_1} D_{i_5}D_{i_6)}\bar{T}_{\alpha k_2} 
+20 \partial_{( i_1} \partial_{i_2} \bar{\Gamma}^{\alpha}_{i_3 k_1} D_{i_4} \cdots D_{i_6)}\bar{T}_{\alpha k_2} \non \\
&+& 14\partial_{( i_1} \bar{\Gamma}^{\alpha}_{i_2 k_1} D_{i_3} \cdots D_{i_6)}\bar{T}_{\alpha k_2} \non \\
&+& 10 \partial_{( i_1} \cdots \partial_{i_3} \bar{\Gamma}^{\alpha}_{i_4 k_1} \left(\partial_{i_5} 
\bar{\Gamma}^{\rho}_{i_6) \alpha} \bar{T}_{\rho k_2} + \alpha \leftrightarrow k_2 \right) \non \\
&+& 10 \partial_{( i_1} \partial_{i_2} \bar{\Gamma}^{\alpha}_{i_3 k_1} \left(\partial_{i_4} \partial_{i_5} 
\bar{\Gamma}^{\rho}_{i_6) \alpha} \bar{T}_{\rho k_2} + \alpha \leftrightarrow k_2 \right) \non \\
&+& 36\partial_{( i_1} \partial_{i_2} \bar{\Gamma}^{\alpha}_{i_3 k_1} \left(\partial_{i_4} 
\bar{\Gamma}^{\rho}_{i_5 \alpha} D_{i_6)}\bar{T}_{\rho k_2} + \alpha \leftrightarrow k_2 \right) \non \\
&+& 5 \partial_{( i_1} \bar{\Gamma}^{\alpha}_{i_2 k_1} \left(\partial_{i_3} \cdots \partial_{i_5}
\bar{\Gamma}^{\rho}_{i_6) \alpha} \bar{T}_{\rho k_2} + \alpha \leftrightarrow k_2 \right) \non \\
&+& 24 \partial_{( i_1} \bar{\Gamma}^{\alpha}_{i_2 k_1} \left(\partial_{i_3} \partial_{i_4} 
\bar{\Gamma}^{\rho}_{i_5 \alpha}D_{i_6)}\bar{T}_{\rho k_2} + \alpha \leftrightarrow k_2 \right) \non \\
&+& 45 \partial_{( i_1} \bar{\Gamma}^{\alpha}_{i_2 k_1} \left(\partial_{i_3} \bar{\Gamma}^{\rho}_{i_4 \alpha}
D_{i_5} D_{i_6)}\bar{T}_{\rho k_2} + \alpha \leftrightarrow k_2 \right) \non \\ 
&+& 15 \left[ \partial_{(i_1} \bar{\Gamma}^{\alpha}_{i_2 k_1} \partial_{i_3} \bar{\Gamma}^{\rho}_{i_4 \alpha}\left( 
\partial_{i_5} \bar{\Gamma}^{\sigma}_{i_6) \rho} \bar{T}_{\sigma k_2} + \rho \leftrightarrow k_2 \right) 
+ \alpha \leftrightarrow k_2 \right] \non \\
&+& \partial_{(i_1} \bar{\Gamma}^{\alpha}_{i_2 k_1} D_{i_3} \cdots D_{i_6)} \bar{T}_{\alpha k_2}
+ k_1 \leftrightarrow k_2 \Big].
\label{coedn}
\feqr

\noi If the second rank tensor with components $\bar{T}_{k_1 k_2}$ is replaced by the metric components 
$\bar{g}_{k_1 k_2}$ then the related covariant derivatives (provided we deal with a torsion free affine connection) 
vanish and the above expressions are simplified. One could derive for $n=5$ the result:
\beqr
\partial_{(i_1} \cdots \partial_{i_5)}\bar{g}_{k_1 k_2} = \frac{4}{3} \left[D_{i_1} \cdots D_{i_3} 
\bar{R}_{k_1 i_4 i_5 k_2} 
+ 2 \left(D_{i_1} \bar{R}_{k_1 i_2 i_3 \rho} \bar{R}^{\rho}_{i_4 i_5 k_2} +  k_1 \leftrightarrow k_2 \right) 
\right].
\label{odty}
\feqr

\noi On the other hand for $n=6$ one gets:
\beqr
\partial_{( i_1} \partial_{i_2} \partial_{i_3} \partial{i_4} \partial_{i_5} \partial_{i_6)}\bar{g}_{k_1 k_2} = & &
\frac{10}{7}D_{(i_1} \cdots D_{i_4} \bar{R}_{k_1 i_5 i_6) k_2}  \non \\
&+& \frac{34}{7}\left(D_{(i_1}D_{i_2} \bar{R}_{k_1 i_3 i_4 \rho} 
\bar{R}^{\rho}_{i_5 i_6) k_2} + k_1 \leftrightarrow k_2 \right) \non \\
&+& \frac{55}{7} D_{(i_1} \bar{R}_{k_1 i_2 i_3 \rho} D_{i_4} \bar{R}^{\rho}_{i_5 i_6) k_2} +
\frac{16}{7} \bar{R}_{k_1 (i_1 i_2 \rho} \bar{R}^{\rho}_{i_3 i_4 l} \bar{R}^{l}_{i_5 i_6) k_2}.
\label{evety}
\feqr

\noi Thus, plugging into\rf{exp} expressions\rf{odty} and\rf{evety} we end up with the following expansion of the 
metric tensor in RNC:
\beqr
g_{k_1 k_2} = & & \bar{g}_{k_1 k_2} + \frac{1}{2!} \frac{2}{3}
\bar{R}_{k_1 i_1 i_2 k_2} \xi^{i_1} \xi^{i_2} \non \\
&+& \frac{1}{3!} D_{i_1} \bar{R}_{k_1 i_2 i_3 k_2} \xi^{i_1} \cdots \xi^{i_3} \non \\
&+& \frac{1}{4!}\frac{6}{5} \left[D_{i_1}D_{i_2} \bar{R}_{k_1 i_3 i_4 k_2} 
+ \frac{8}{9} \bar{R}_{k_1 i_1 i_2 m} \bar{R}^{m}_{i_3 i_4 k_2}\right] \xi^{i_1} \cdots \xi^{i_4}\non \\
&+& \frac{1}{5!} \frac{4}{3} \left[D_{i_1} \cdots D_{i_3} \bar{R}_{k_1 i_4 i_5 k_2} 
+ 2 \left(D_{i_1} \bar{R}_{k_1 i_2 i_3 \rho} \bar{R}^{\rho}_{i_4 i_5 k_2} +  k_1 \leftrightarrow k_2 \right) 
\right] \xi^{i_1} \cdots \xi^{i_5} \non \\
&+& \frac{1}{6!} \frac{10}{7}\Big[ D_{i_1} \cdots D_{i_4} \bar{R}_{k_1 i_5 i_6 k_2} + \frac{17}{5}\left(D_{i_1}D_{i_2} 
\bar{R}_{k_1 i_3 i_4 \rho} 
\bar{R}^{\rho}_{i_5 i_6 k_2} + k_1 \leftrightarrow k_2 \right) \non \\
&+& \frac{11}{2}D_{i_1} \bar{R}_{k_1 i_2 i_3 \rho} D_{i_4} \bar{R}^{\rho}_{i_5 i_6 k_2} +
\frac{8}{5}\bar{R}_{k_1 i_1 i_2 \rho} \bar{R}^{\rho}_{i_3 i_4 l} \bar{R}^{l}_{i_5 i_6 k_2} \Big] 
\xi^{i_1} \cdots \xi^{i_6}\non \\
&+& O(\xi^{i_1} \cdots \xi^{i_7}).
\label{exme}
\feqr

\noi The expression\rf{exme} will play a crucial role in calculating the contribution steming from  the quadratic 
Christoffel symbol term in the interaction Lagrangian. It is also in perfect agreement with the closed formula 
for these coefficients which are encoded in the integral representation of \cite{schubert}.
%%%%%%%%%%%%%%%%%%%%%%%%%%%%%%%%%%%%%%%%%%%%%%%%%%%%%%%%%%%%%%%%%%%%%%%%%%%%%%%%%%%%%%%%%%%%%%%%%%%%%%%%%%%%%%%%%%%%%%%%%%%

						%%%%%%%%%%%%%%%%
						%   SECTION 2  %
						%%%%%%%%%%%%%%%%

\section{The integrated trace anomaly}

Anomalies in (even) n-dimensional quantum field theories\footnote{ For a manifold having odd dimensionality one cannot form 
a scalar out of an odd number of derivatives.} are expressed in the Fujikawa \cite{kazuo} approach as
\footnote{We perform the usual redefinition of the scalar fields $\tilde{\phi}=g^{\frac{1}{4}}\phi$ which leaves
 the Jacobian invariant under the similarity transformation $g^{\frac{1}{4}}$.}:
\beqr
An_{W}=\lim_{\beta \rightarrow 0}Tr \left( \mathcal{J} e^{-\frac{\beta}{\hbar} \hat{\mathcal{R}}} \right)
\label{fuji} 
\feqr

\noi where $\mathcal{J}$ is the Jacobian $\frac{\partial\delta \tilde{\phi}(x)}{\partial 
\tilde{\phi}(y)}= f(x) \tilde{\phi}$ of the 
fields $\tilde{\phi}(y)$
and the regulator $\hat{\mathcal{R}}$ for consistent anomalies is uniquely determined 
\cite{troost}. For local Weyl anomalies and for real scalar fields one finds that:
\beqr
\hat{\mathcal{R}} &=& \hat{H} - \frac{1}{2} \hbar^2 \xi R(\hat{x}) \non \\
&=& \frac{1}{2}g^{-\frac{1}{4}}(\hat{x})\hat{p}_{i}g^{ij}(\hat{x})g^{\frac{1}{2}}(\hat{x})\hat{p}_{j}
g^{-\frac{1}{4}}(\hat{x}) - \frac{1}{2} \hbar^2 \xi R(\hat{x}); \quad g(\hat{x})=detg_{ij}(\hat{x}).
\label{regul}
\feqr
     
\noi The second term in \rf{regul} is the well-known improvement potential term and the dimensionless coefficient 
$\xi=\frac{(n-2)}{n(n-1)}$ takes the value $\xi(n=6)=\frac{2}{15}$ in the present case. 
The Hamiltonian is Einstein invariant which means that $\hat{H}$ commutes with the generator
$\hat{G}(\xi)=\frac{1}{2i\hbar}(\hat{p}_{i}\xi^{i}(\hat{x})+\xi^{i}(\hat{x})\hat{p}_{i})$ 
of the infinitesimal target space diffeomorphisms $\hat{x}_{i} \rightarrow \hat{x}_{i} + \xi_{i}(\hat{x})$.

Consider a spin-0 field which lives on an n-dimensional compact Riemaniann manifold equipped with its standard 
(metric-compatible, torsion-free) connection and having no boundary. 
In addition decompose the paths $x^{i}(\sigma)$ into a 
constant part $x_{0}^{i}$ satisfying the free field equations and a quantum fluctuating $q^{i}(\sigma)$ one 
vanishing at the time boundaries. The Weyl anomaly for a real 
spin-0 field may represented by the Euclidean quantum mechanical path integral \cite{takis}:
\beqr
An_{W}(s=0, n) &=& \lim_{\beta \rightarrow 0}Tr \left(f(x) e^{-\beta \hat{H}} \right) \non\\
&=& \lim_{\beta \rightarrow 0} \left( \frac{1}{(2\pi \beta \hbar)^{n/2}}
\int dx^{i}_{0}\prod_{i=1}^{n} \sqrt{g(x^{i}_{0})} f(x_{0}^{i}) <e^{-\frac{1}{\hbar}S^{int}}> \right)
\label{tracean}
\feqr

\noi where $f(x)$ is an arbitrary function, 
$-\frac{1}{\hbar}S^{int}=S^{int}_{1}+S^{int}_{2}$ and 
\footnote{The expectation value $<\cdots>$ means that all quantum fields must be contracted using 
the appropriate propagators. 
A bar over the various geometrical quantities indicates that they depend exclusively on $x^{i}_{0}$.}:
\beqr
S^{in}_{1}&=&\frac{1}{2\beta \hbar}\int_{-1}^{0} [g_{ij}(x_{0}+q)-g_{ij}(x_0)] 
\left(\dot{q}^{i}\dot{q}^{j}+ b^{i}c^{j}+ a^{i}a^{j} \right) d\sigma \non \\
&=& \frac{1}{\beta \hbar}\int_{-1}^{0} \bigg[\frac{1}{6} \bar{R}_{i i_1 i_2 j}q^{i_1}q^{i_2}+
\frac{1}{12}D_{i_1}\bar{R}_{i i_2 i_3 j} q^{i_1}q^{i_2}q^{i_3} \non \\
&+&\frac{3}{5!}\bigg(D_{i_1}D_{i_2}\bar{R}_{i i_3 i_4 j}+\frac{8}{9}\bar{R}_{i i_1 i_2 k}
\bar{R}^{k}_{\hspace{0.05in}i_3 i_4 j} \bigg) q^{i_1} \cdots q^{i_4} \non \\ 
&+& \frac{4}{6!} \bigg(D_{i_1}\cdots D_{i_3} \bar{R}_{i i_4 i_5 j}+ 2(D_{i_1}\bar{R}_{i i_2 i_3 k} 
\bar{R}^{k}_{\hspace{0.05in} i_4 i_5 j} + i \leftrightarrow j) \bigg) q^{i_1} \cdots q^{i_5} \non \\
&+& \frac{5}{7!}\bigg( D_{i_1} \cdots D_{i_4} \bar{R}_{i i_5 i_6 j} + \frac{17}{5}\left(D_{i_1}D_{i_2} 
\bar{R}_{i i_3 i_4 k} \bar{R}^{k}_{\hspace{0.05in}i_5 i_6 j} + i \leftrightarrow j \right)+\frac{11}{2}
D_{i_1}\bar{R}_{i  i_2 i_3 k} D_{i_4} \bar{R}^{k}_{\hspace{0.05in}i_5 i_6 j} \non \\
&+& \frac{8}{5}\bar{R}_{i i_1 i_2 k} \bar{R}^{k}_{\hspace{0.05in}i_3 i_4 l} \bar{R}^{l}_{\hspace{0.05in}i_5 i_6 j} \bigg)
 q^{i_1} \cdots q^{i_6} + O(q^{i_1} \cdots q^{i_7})\bigg] 
\left(\dot{q}^{i}\dot{q}^{j}+ b^{i}c^{j}+ a^{i}a^{j} \right) d\sigma
\label{int1}
\feqr

\beqr
S^{int}_{2} &=&-\beta \hbar \frac{1}{8} \int_{-1}^{0} \bigg[(1-4\xi)R+ g^{ij} \Gamma^{l}_{\hspace{0.05in}k i}
\Gamma^{k}_{\hspace{0.05in}l j} \bigg](x_0 +q) d\tau \non \\
&=& -\beta \hbar \frac{1}{8} \int_{-1}^{0} \bigg[ (1-4\xi) \bigg( \bar{R}+ q^{i_1}D_{i_1}\bar{R} +
\frac{1}{2!} q^{i_1}q^{i_2}D_{i_1}D_{i_2}\bar{R} + \frac{1}{3!}q^{i_1}\cdots q^{i_3}D_{i_1}\cdots D_{i_3}\bar{R} \non \\
&+& \frac{1}{4!}q^{i_1}\cdots q^{i_4}D_{i_1}\cdots D_{i_4}\bar{R} \bigg) + \bar{g}^{ij} \partial_{i_1}
\bar{\Gamma}^{l}_{\hspace{0.05in}i k}\partial_{i_2}\bar{\Gamma}^{k}_{\hspace{0.05in}l j} q^{i_1} q^{i_2} \non \\
&+& \frac{1}{2!}\bar{g}^{ij} \left(\partial_{i_1}\partial_{i_2}\bar{\Gamma}^{l}_{\hspace{0.05in}i k}\partial_{i_3}
\bar{\Gamma}^{k}_{\hspace{0.05in}l j}+
\partial_{i_1}\bar{\Gamma}^{l}_{\hspace{0.05in}i k}\partial_{i_2}\partial_{i_3}\bar{\Gamma}^{k}_{\hspace{0.05in}l j}\right) 
q^{i_1} \cdots q^{i_3} \non \\
&-& \bigg( \frac{1}{3}\bar{R}^{i \hspace{0.2in} j}_{\hspace{0.05in}i_1 i_2}\partial_{i_3}\bar{\Gamma}^{l}_{\hspace{0.05in}i k}
\partial_{i_4}\bar{\Gamma}^{k}_{\hspace{0.05in}l j}  - \frac{1}{4} \bar{g}^{ij}
\partial_{i_1}\partial_{i_2} \bar{\Gamma}^{l}_{\hspace{0.05in}i k}\partial_{i_3}\partial_{i_4} 
\bar{\Gamma}^{k}_{\hspace{0.05in}l j} \non \\
&-& \frac{\bar{g}^{ij}}{3!}\left(\partial_{i_1} \cdots \partial_{i_3} \bar{\Gamma}^{l}_{\hspace{0.05in}i k} \partial_{i_4}
\bar{\Gamma}^{k}_{\hspace{0.05in}l j} + \partial_{i_1} \bar{\Gamma}^{l}_{\hspace{0.05in}i k}\partial_{i_2} \cdots \partial_{i_4}
\bar{\Gamma}^{k}_{\hspace{0.05in}l j} \right)
\bigg) q^{i_1} \cdots q^{i_4}\non \\ 
&+& O(q^{i_1} \cdots q^{i_5}) \bigg] d\tau .
\label{int2}
\feqr

\noi As emphasized in \cite{kostas}, the above expression is the continuum limit of a rigorous discrete result.  

In the action \rf{int1} $(b^{i}, c^{j})$ and $\alpha^{i}$ is a set of anticommuting and commuting Lee-Yang ghosts 
respectively \cite{peter}. Their existence is imposed by the integration over the momenta 
$p^{i}(\tau)$ in the transition
from phase space to configuration space. A measure factor $\sqrt{g}$ is then produced at each point of the path 
and by exponentiating it, introducing the 
Lee-Yang ghosts, we are led to a perfectly regular term in the action. The existence of ghost fields 
also removes ultraviolet divergencies at higher loops and as a consequence all integrals are finite.

The non-covariant $\hbar^2$ terms, $R$ and $\Gamma \Gamma$, which are essential for the general coordinate invariance of the 
transition element are created \cite{kostas} by Weyl ordering the Hamiltonian $\hat{H}$ of \rf{regul}. The contribution of 
these terms to the trace anomaly is found by first Taylor expanding them in RNC and then substituting 
the partial derivatives of the Christoffel symbols by the polynomials of $R$ and $DR$:
\beqr
\partial_{i_1}\bar{\Gamma}^{l}_{\hspace{0.05in}(i_2  k)} &=&  -\frac{2}{3}
\bar{R}^{l}_{\hspace{0.05in}(i_2 k) i_1} \\
\partial_{(i_1}\partial_{i_2}\bar{\Gamma}^{l}_{\hspace{0.05in}i_3) k} &= &-\frac{1}{2}D_{(i_1}
\bar{R}^{l}_{\hspace{0.05in}i_2 k i_3 )} \\
\partial_{(i_1}\partial_{i_2} \partial_{i_3}
\bar{\Gamma}^{l}_{\hspace{0.05in}i_4) k} &= &-\frac{3}{5} \left[D_{(i_1} D_{i_2} \bar{R}^{l}_{\hspace{0.05in}i_3 k i_4 )}
+ \frac{2}{9} \bar{R}^{l}_{\hspace{0.05in}(i_1 i_2 \alpha} \bar{R}^{\alpha}_{\hspace{0.05in}i_3 i_4 )k}
\right]. 
\label{id1} 
\feqr

\noi The symmetrization of the various indices in the above identities is understood with the inclusion of a 
$\frac{1}{n!}$ factor. All other extra terms in \rf{int1} and \rf{int2} are produced by expanding the metric in 
Riemann normal coordinates with the help of \rf{exme}.

Propagators are derived in closed form through the discretised configuration space path integrals via a midpoint rule. 
In this way ambiguities arising from products of 
distributions are resolved. Taking the continuum limits of the propagators one can read off the Feynman rules. One then finds that 
$\delta(\sigma -\tau)$ is to be considered as a Kronecker delta $\delta_{ij}$ and the propagators depend on the discrete 
Heaviside function $\theta_{ii}=\frac{1}{2}$. 
The continous two point Green function may also be determined by the equation:
\beqr
\frac{\partial^2}{\partial \sigma^2} \Delta(\sigma -\tau)=\delta(\sigma -\tau)
\label{green} 
\feqr

\noi subjected to the boundary conditions on the interval $[-1,0]$ (end-point approach):
\beqr
\Delta(0,\tau)=\Delta(-1,\tau)=\Delta(\sigma,0)=\Delta(\sigma,-1)=0.
\label{bound}
\feqr

\noi The Feynman propagator is then formally found to be:
\beqr
\Delta(\sigma,\tau)=\sigma(\tau +1)+(\tau -\sigma) \theta (\tau - \sigma).
\label{feypr}
\feqr 
The propagators of the various fields are proportional to $\beta \hbar$ and given by:
\beqr
q^{i}(\sigma)q^{j}(\tau)&=& -\beta \hbar \bar{g}^{ij} \Delta(\sigma,\tau) \\
q^{i}(\sigma)\dot{q}^{j}(\tau)&=&-\beta \hbar \bar{g}^{ij} [\sigma + \theta(\tau-\sigma)] 
=-\beta \hbar \bar{g}^{ij} \Delta^{\bullet}(\sigma,\tau) \\
\dot{q}^{i}(\sigma)q^{j}(\tau)&=&-\beta \hbar \bar{g}^{ij} [\tau + \theta(\sigma -\tau)] 
=-\beta \hbar \bar{g}^{ij} \, {}^{\bullet}\Delta(\sigma,\tau) \\
\dot{q}^{i}(\sigma)\dot{q}^{j}(\tau)&=&-\beta \hbar \bar{g}^{ij}[1-\delta(\sigma -\tau)] 
=-\beta \hbar \bar{g}^{ij} \, {}^{\bullet}\Delta^{\bullet}(\sigma,\tau) \\
b^{i}(\sigma)c^{j}(\tau)&=&-2\beta \hbar \bar{g}^{ij} \delta(\sigma -\tau)   \\
a^{i}(\sigma)a^{j}(\tau)&=&+\beta \hbar \bar{g}^{ij} \delta(\sigma - \tau).
\label{propa}
\feqr 

\noi All other Wick contractions of the fields vanish. Regarding the vertices they may be read directly from the 
continuum $S_{int}$ given by \rf{int1} and \rf{int2}.

%%%%%%%%%%%%%%%%%%%%%%%%%%%%%%%%%%%%%%%%%%%%%%%%%%%%%%%%%%%%%%%%%%%%%%%%%%%%%%%%%%%%%%%%%%%%%%%%%%%%%%%%%%%%%%%%%%%%%%%%%%%

                                               %%%%%%%%%%%%%%%%%%%%%%%%
                                               % THE FEYNMAN DIAGRAMS %  
                                               %%%%%%%%%%%%%%%%%%%%%%%%

%%%%%%%%%%%%%%%%%%%%%%%%%%%%%%%%%%%%%%%%%%%%%%%%%%%%%%%%%%%%%%%%%%%%%%%%%%%%%%%%%%%%%%%%%%%%%%%%%%%%%%%%%%%%%%%%%

\section{The Feynman diagrams and the associated contributions}

Perturbative expansion of $<e^{-\frac{1}{\hbar} S^{int}}>$, keeping only terms that cancel the $(\beta \hbar)^{-3}$ 
factor in the measure of the trace anomaly, provides us with the following distinct interactions as well as 
connected and disconnected diagrams having at most four loops
\footnote{The recipee the diagrams have been drawn is: A line closing on a single vertex is represented by a Green function 
$G(\sigma,\sigma)$ with one integration variable while a line connecting two vertices stands for $G(\sigma,\tau)$ 
and indicates two integration variables.}:

\begin{enumerate}
\item 
%%\color{red}
\beqr
I_1 &=&\frac{1}{\beta \hbar} \frac{5}{7!} <\int_{-1}^{0} \bigg[ \bigg[ D_{i_1} \cdots 
D_{i_4} \bar{R}_{i i_5 i_6 j} + \frac{17}{5}\left(D_{i_1}D_{i_2} 
\bar{R}_{i i_3 i_4 k} \bar{R}^{k}_{\hspace{0.05in}i_5 i_6 j} + i \leftrightarrow j \right) \non\\
&+&\frac{11}{2}D_{i_1}\bar{R}_{i  i_2 i_3 k} D_{i_4} \bar{R}^{k}_{\hspace{0.05in} i_5 i_6 j} \non \\
&+& \frac{8}{5}\bar{R}_{i i_1 i_2 k} \bar{R}^{k}_{\hspace{0.05in} i_3 i_4 l} \bar{R}^{l}_{\hspace{0.05in} i_5 i_6 j} \bigg]
 q^{i_1} \cdots q^{i_6}  
\left(\dot{q}^{i}\dot{q}^{j}+ b^{i}c^{j}+ a^{i}a^{j} \right) \bigg](\sigma) d\sigma > 
\label{vertex1}
\feqr

%%\color{blue}
\vspace{0.5in}
\begin{picture}(0,0)
%%\color{green}
%%\graphpaper(-10,-160)(450,180)
%%\color{blue}
\put(28,4){\usebox{\ninety}}
\put(23,0){\usebox{\onehun}}
\put(28,-5){\usebox{\twosev}}
\put(32,0){\usebox{\threehun}}
\put(30,0){\circle*{4}}
\put(70,-3){$= \frac{5}{7!} \left(\beta \hbar \right)^3 \bigg\{ \frac{1}{120}\bigg( \Box^2 \bar{R} 
+2D_{a} \Box D^{a} \bar{R} \bigg)+ \frac{1}{60}\bigg( \Box D_{a}D_{b}\bar{R}^{ab} + D_{a}D_{b}\Box \bar{R}^{ab}$}
\put(70,-33){$+D_{a}D_{b}D^{a}D_{c}\bar{R}^{bc}+ D_{a}\Box D_{b}\bar{R}^{ab}+ 
D_{a}D_{b}D_{c}D^{b}\bar{R}^{ac} + D_{a}D_{b}D_{c}D^{a}\bar{R}^{bc}\bigg)$}
\put(-10,-63){$-\frac{1}{120}\frac{34}{5}\bigg[\bar{R}^{ab} \Box \bar{R}_{ab} 
+ \frac{3}{2} \bar{R}^{abcd} \Box \bar{R}_{abcd} + 2\bar{R}^{ab} D^{c}D^{d} \bar{R}_{dacb} 
+ 2\bar{R}^{a b c d} D_{d}D_{b} \bar{R}_{ac}$}
\put(-10,-93){$+12\bar{R}^{a b c d} D_{(e}D_{d)} \bar{R}^{e}_{\hspace{0.1in} abc} \bigg]
-\frac{1}{120} \frac{11}{2} \bigg[ \left(D_{a}\bar{R}_{bc} \right)^2 + 4 D_{a} \bar{R}_{bc} 
D^{d} \bar{R}_{dcab} + \frac{3}{2} \left(D_{a} \bar{R}_{bcde}\right)^2$}
\put(-10,-123){$+ 3D_{a} \bar{R}^{abcd} D^{e} \bar{R}_{ebcd}
+ 3 D^{a} \bar{R}^{bcde} D_{e} \bar{R}_{bcda}\bigg] +
\frac{1}{120} \frac{8}{5} \bigg[\bar{R}_{ab} \bar{R}^{b}_{c}\bar{R}^{ac} + 
9\bar{R}^{ab}\bar{R}_{a}^{\hspace{0.1in} edc}\bar{R}_{bcde} $}
\put(-10, -153){$-\bar{R}_{abcd}\bar{R}^{\hspace{0.1in} a \hspace{0.1in} c}_{e \hspace{0.1in} f} \bar{R}_{bedf}
+\frac{7}{2} \bar{R}_{abcd}\bar{R}_{ef}^{\hspace{0.2in} ab} \bar{R}^{efcd} \bigg] \bigg\}$}
\end{picture}

\vspace{2.5in}

\item 
%%\color{red}
\beqr
I_2 &=& - \beta \hbar \frac{1}{8} < \int_{-1}^{0} \bigg[ \bigg[\frac{1}{4!}(1-4\xi) D_{i_1} \cdots D_{i_4}\bar{R}
-  \frac{1}{3}\bar{R}^{i \hspace{0.2in}j}_{\hspace{0.05in} i_1 i_2}\partial_{i_3}
\bar{\Gamma}^{l}_{\hspace{0.05in} i k}
\partial_{i_4}\bar{\Gamma}^{k}_{\hspace{0.05in}l j}  \non \\
&+& \frac{1}{4} \bar{g}^{ij}
\partial_{i_1}\partial_{i_2}\Gamma^{l}_{\hspace{0.05in}i k}\partial_{i_3}\partial_{i_4}
\Gamma^{k}_{\hspace{0.05in}l j} \non \\
&+& \frac{\bar{g}^{ij}}{3!}\left(\partial_{i_1} \cdots \partial_{i_3} \Gamma^{l}_{\hspace{0.05in}i k} \partial_{i_4}
\Gamma^{k}_{\hspace{0.05in}l j} + \partial_{i_1}\Gamma^{l}_{\hspace{0.05in}i k}\partial_{i_2} \cdots \partial_{i_4}
\Gamma^{k}_{\hspace{0.05in}l j} \right)
\bigg] q^{i_1} \cdots q^{i_4} \bigg](\sigma) d\sigma >
\label{vertex2}
\feqr

%%\color{blue}
\vspace{0.1in}

\begin{picture}(0,0)
\thicklines
%%\color{green}
%%\graphpaper(-10,-220)(420,230)
%%\color{blue}
\put(30,0){\circle{20}}
\put(40,0){\circle*{3}}
\put(50,0){\circle{20}}
\put(70,-3){$=- \left(\beta \hbar\right)^3 \frac{1}{8}\bigg[\frac{1}{720}(1-4\xi) \Bigg( 3 \Box^2 \bar{R}
-2D_{a}\bar{R}^{ab} D_{b} \bar{R} -2\bar{R}^{ab}D_{a}D_{b}\bar{R} \Bigg)$}
\put(70,-33){$ - \frac{1}{540}
\bigg(\frac{1}{2}\bar{R}^{ab}\bar{R}_{a}^{\hspace{0.1in} cde}\bar{R}_{bcde} + \frac{1}{3}\bar{R}^{abcd}
\bar{R}_{a \hspace{0.1in} c}^{\hspace{0.1in} e \hspace{0.1in} f}\bar{R}_{bedf}+\frac{13}{12}\bar{R}^{abcd}
\bar{R}_{ab}^{\hspace{0.1in} ef} \bar{R}_{cdef} \bigg) $}
\put(70,-63){$-\frac{1}{1728}D^{a}\bar{R}^{bc}D_{d}\bar{R}_{ab \hspace{0.1in} c}^{\hspace{0.1in}d}+
\frac{1}{17280} \left(D^{a}\bar{R}^{bc}\right)^2-\frac{5}{2304}D_{a}\bar{R}^{abcd}D_{e}\bar{R}^{e}_{\hspace{0.1in}bcd}$}
\put(70,-93){$ +\frac{1}{5760}D^{a}\bar{R}^{bc}D_{c}\bar{R}_{ab} 
-\frac{1}{480}\left(D^{a}\bar{R}^{bcde}\right)^2
 -\frac{1}{360}D^{a}\bar{R}^{bcde}D_{e}\bar{R}_{bcda}$} 
\put(70,-123){$+\frac{1259}{518400}\bar{R}^{abcd}\bar{R}_{ab}^{\hspace{0.1in} ef}\bar{R}_{cdef}+ 
\frac{47}{86400}\bar{R}^{ab} \bar{R}_{a}^{\hspace{0.1in} cde} \bar{R}_{bcde} 
-\frac{1}{7200}\bar{R}^{abcd}D_{d}D_{b}\bar{R}_{ac} $}
\put(70,-153){$ -\frac{23}{4800}\bar{R}^{abcd}D_{d}D_{e}\bar{R}^{e}_{\hspace{0.1in} abc} 
-\frac{7}{3200} \bar{R}^{abcd} \Box \bar{R}_{abcd}
-\frac{7}{1600}\bar{R}^{abcd}D_{e}D_{d}\bar{R}^{e}_{\hspace{0.1in} abc}$}
\put(70,-183){$ +\frac{209}{129600}\bar{R}^{abcd}
\bar{R}_{a \hspace{0.1in}c }^{\hspace{0.1in}e \hspace{0.1in} f} \bar{R}_{bedf} \bigg] $}
\end{picture}

\vspace{2.4in}

\item 
%%\color{red}
\beqr
I_3 &=& \frac{1}{2} \left(\frac{1}{\beta \hbar}\right)^2 <\int_{-1}^{0}\int_{-1}^{0} \bigg[ 
\left(\frac{1}{12}\right)^2 \bigg[ D_{i_1}\bar{R}_{i i_2 i_3 j} q^{i_1} \cdots q^{i_3}
\left(\dot{q}^{i}\dot{q}^{j}+ b^{i}c^{j}+ a^{i}a^{j} \right) \bigg](\sigma) \non \\
&\times& \!\!\! \bigg[ D_{i_4}\bar{R}_{k i_5 i_6 l} q^{i_4} \cdots q^{i_6}
\left(\dot{q}^{k}\dot{q}^{l}+ b^{k}c^{l}+ a^{k}a^{l} \right) \bigg](\tau) \non \\
&+& \!\!\! \frac{1}{5!} \bigg[ \bar{R}_{i i_1 i_2 j} q^{i_1}q^{i_2} 
\left(\dot{q}^{i}\dot{q}^{j}+ b^{i}c^{j}+ a^{i}a^{j} \right)\bigg](\sigma)  \non \\
&\times&  \bigg[ \bigg(D_{i_3}D_{i_4}\bar{R}_{k i_5 i_6 l}+\frac{8}{9}\bar{R}_{k i_3 i_4 m}
\bar{R}^{m}_{\hspace{0.05in} i_5 i_6 l} \bigg)q^{i_3} \cdots q^{i_6}
\left(\dot{q}^{k}\dot{q}^{l}+ b^{k}c^{l}+ a^{k}a^{l} \right)\bigg](\tau)\bigg] \non \\ 
&& d\sigma d\tau > 
\label{vertex3}
\feqr

\vspace{0.1in}

%%\color{blue}
\begin{picture}(0,0)
%%\color{green}
%%\graphpaper(-10,-180)(500,200)
%%\color{blue}
\thicklines
%%%%%%%%%%%%%%%%%%%%%%%%%%%%%%%%%%%%
\put(30,0){\circle{20}}
\put(20,0){\circle*{3}}
\put(40,0){\circle*{3}}
\qbezier(20,0)(30,8)(40,0)
\qbezier(20,0)(30,-8)(40,0)
\put(20,0){\line(1,0){20}}
%%%%%%%%%%%%%%%%%%%%%%%%%%%%%%%%%%%%
\put(45,-3){$+$}
%%%%%%%%%%%%%%%%%%%%%%%%%%%%%%%%%%%
\put(68,4){\usebox{\ninety}}
\put(68,-5){\usebox{\twosev}}
\put(70,0){\circle*{4}}
\put(70,0){\line(1,0){30}}
\put(98,4){\usebox{\ninety}}
\put(98,-5){\usebox{\twosev}}
\put(100,0){\circle*{4}}
%%%%%%%%%%%%%%%%%%%%%%%%%%%%%%%%%%%%%%%
\put(115,-3){$+$}
%%%%%%%%%%%%%%%%%%%%%%%%%%%%%%%%%%%%%%%
\put(140,0){\circle{20}}
\put(150,0){\circle*{3}}
\put(160,0){\circle{20}}
\put(170,0){\circle*{3}}
\put(150,0){\line(1,0){20}}
\put(180,0){\circle{20}}
%%%%%%%%%%%%%%%%%%%%%%%%%%%%%%%%%%%%%%%%
\put(195,-3){$+$}
%%%%%%%%%%%%%%%%%%%%%%%%%%%%%%%%%%%%%%%
\put(207,-3){$\bigg[$}
\put(230,0){\circle{20}}
\put(240,0){\circle*{3}}
\put(250,0){\circle{20}}
%%%%%%%%%%%%%%%%%%%%%%%%%%%%%%%%%%%%
\put(265,-3){$\times$}
%%%%%%%%%%%%%%%%%%%%%%%%%%%%%%%%%%%%
\put(298,4){\usebox{\ninety}}
\put(295,-3){\usebox{\twotwofive}}
\put(301,-3){\usebox{\thonefive}}
\put(300,0){\circle*{3}}
%%%%%%%%%%%%%%%%%%%%%%%%%%%%%%%%%%%%%
\put(328,-3){$\bigg]$}
\put(335,-3){$+$}
%%%%%%%%%%%%%%%%%%%%%%%%%%%%%%%%%%%%%%
\put(360,0){\circle{20}}
\put(370,0){\circle*{3}}
\put(380,0){\circle{20}}
\put(391,2){\usebox{\fourtyfive}}
\put(391,-3){\usebox{\thonefive}}
\put(390,0){\circle*{3}}
%%%%%%%%%%%%%%%%%%%%%%%%%%%%%%%%%%%%%
\put(415,-3){$+$}
%%%%%%%%%%%%%%%%%%%%%%%%%%%%%%%%%%%%
\put(440,0){\circle{20}}
\put(430,0){\circle*{3}}
\put(450,0){\circle*{3}}
\qbezier(430,0)(440,8)(450,0)
\qbezier(430,0)(440,-8)(450,0)
\put(460,0){\circle{20}}
\put(30,-53){$= -\frac{1}{2} \frac{1}{144}  \left(\beta \hbar \right)^3 
\bigg[\frac{1}{10} D_{a}\bar{R}_{bcde}D^{e} \bar{R}^{bcda}+\frac{1}{10}D_{a}\bar{R}^{abcd}D^{e} \bar{R}_{ebcd}
+\frac{1}{20} \left(D_{a}\bar{R}_{bcde}\right)^2$}
\put(30,-83){$ -\frac{2}{15}D_{a}\bar{R}_{bc}D_{d}\bar{R}^{abcd} +\frac{1}{30} \left( D_{a}\bar{R}_{bc}\right)^2 -
\frac{1}{48}D_{a}\bar{R}^{ab} D_{b}\bar{R} -\frac{1}{48}D_{a}\bar{R}^{ab} D^{c}\bar{R}_{cb} -
\frac{1}{192} \left( D_{a}\bar{R}\right)^2$}
\put(30,-113){$+\frac{1}{5!}\bigg(\frac{108}{15} \bar{R}_{abcd}\Box \bar{R}^{abcd}
+\frac{72}{15}\bar{R}_{ab}\Box \bar{R}^{ab}
-\frac{3}{2}\bar{R}\Box \bar{R} +\frac{432}{30} D_{a}D^{b}\bar{R}^{acde} \bar{R}_{bcde} +
\frac{432}{30}D^{a}D_{b}\bar{R}^{bcde} \bar{R}_{acde}$}
\put(30,-143){$ +\frac{144}{15}D_{a}D_{b}\bar{R}^{bcad}\bar{R}_{cd}+\frac{144}{15}D_{a}D_{b}\bar{R}_{cd} \bar{R}^{adbc}
-3\bar{R}D_{a}D_{b}\bar{R}^{ab}+\frac{4}{3}\bar{R}\left(\bar{R}_{ab}\right)^2-\frac{192}{5}\bar{R}_{abcd}\bar{R}^{bcde}
\bar{R}^{a}_{e}$}
\put(30,-173){$+\frac{64}{15}\bar{R}_{abcd}\bar{R}^{b \hspace{0.1in} d}_{\hspace{0.1in} e \hspace{0.1in} f }
\bar{R}^{eafc}-\frac{224}{15}\bar{R}_{abcd} \bar{R}^{abef} \bar{R}^{cd}_{\hspace{0.1in} ef}-
\frac{64}{15}\bar{R}_{ab}\bar{R}^{b}_{c}\bar{R}^{ca}+2\bar{R}\left(\bar{R}_{abcd} \right)^2  \bigg) \bigg]$}
\end{picture}
\vspace{3.5in}

\item 
%%\color{red}
\beqr
I_4 &=& -\frac{1}{8} < \int_{-1}^{0}\int_{-1}^{0} \bigg[ \frac{1}{12}(1-4\xi) 
\bigg[  \bar{R}_{i i_1 i_2 j} q^{i_1}q^{i_2} 
\left(\dot{q}^{i}\dot{q}^{j}+ b^{i}c^{j}+ a^{i}a^{j} \right)\bigg](\sigma) \non \\
&\times& \bigg[ D_{i_3}D_{i_4}\bar{R} q^{i_3} q^{i_4}\bigg](\tau) \non \\
&+& \frac{1}{6} \bigg[ \bar{R}_{i i_1 i_2 j} q^{i_1}q^{i_2} 
\left(\dot{q}^{i}\dot{q}^{j}+ b^{i}c^{j}+ a^{i}a^{j} \right) \bigg](\sigma) \non\\
&\times& \bigg[ \bar{g}^{mn} \partial_{i_3}
\bar{\Gamma}^{l}_{\hspace{0.05in}m k}\partial_{i_4}\bar{\Gamma}^{k}_{\hspace{0.05in}l n} 
q^{i_3} q^{i_4}\bigg](\tau) \non \\
&+& \frac{1}{12} (1-4\xi) \bigg[ D_{i_1}\bar{R}_{i i_2 i_3 j} q^{i_1} \cdots q^{i_3}
\left(\dot{q}^{i}\dot{q}^{j}+ b^{i}c^{j}+ a^{i}a^{j} \right) \bigg] (\sigma) \non \\
&\times& \bigg[D_{i_4}\bar{R} q^{i_4} \bigg](\tau) 
+ \frac{3}{5!} (1-4\xi) \bigg[ \bigg( D_{i_1}D_{i_2}\bar{R}_{m i_3 i_4 n}+\frac{8}{9}\bar{R}_{m i_1 i_2 k}
\bar{R}^{k}_{\hspace{0.05in} i_3 i_4 n} \bigg) \non \\
&\times&q^{i_1} \cdots q^{i_4}
\left(\dot{q}^{m}\dot{q}^{n}+ b^{m}c^{n}+ a^{m}a^{n} \right) \bigg](\sigma)  \bar{R} d\sigma d\tau >
\label{vertex4}
\feqr

\vspace{0.3in}

%%\color{blue}
\begin{picture}(0,0)
%%\color{green}
%%\graphpaper(-10,-90)(450,130)
%%\color{blue}
\thicklines
\put(27,-3){$\bigg[$}
\put(50,0){\circle{20}}
\put(60,0){\circle*{3}}
\put(70,0){\circle{20}}
%%%%%%%%%%%%%%%%%%%%%%%%%%%%%%%%%%%%%
\put(85,-3){$\times$}
%%%%%%%%%%%%%%%%%%%%%%%%%%%%%%%%%%%%%%
\put(110,0){\circle{20}}
\put(120,0){\circle*{3}}
%%%%%%%%%%%%%%%%%%%%%%%%%%%%%%%%%%%%%%
\put(128,-3){$\bigg]$}
\put(135,-3){$+$}
%%%%%%%%%%%%%%%%%%%%%%%%%%%%%%%%%%%%%%
\put(160,0){\circle{20}}
\put(170,0){\circle*{3}}
\put(180,0){\circle{20}}
\put(190,0){\circle*{3}}
%%%%%%%%%%%%%%%%%%%%%%%%%%%%%%%%%%%%
\put(195,-3){$+$}
%%%%%%%%%%%%%%%%%%%%%%%%%%%%%%%%%%%%
\put(218,4){\usebox{\ninety}}
\put(218,-3){\usebox{\twosev}}
\put(220,0){\circle*{3}}
\put(220,0){\line(1,0){20}}
\put(240,0){\circle*{3}}
%%%%%%%%%%%%%%%%%%%%%%%%%%%%%%%%%%%%
\put(245,-3){$+$}
\put(257,-3){$\bigg[$}
%%%%%%%%%%%%%%%%%%%%%%%%%%%%%%%%%%%%%
\put(288,4){\usebox{\ninety}}
\put(285,-2){\usebox{\twotwofive}}
\put(291,-3){\usebox{\thonefive}}
\put(290,0){\circle*{3}}
%%%%%%%%%%%%%%%%%%%%%%%%%%%%%%%%%%%%%%
\put(315,-3){$\times$}
\put(330,0){\circle*{3}}
\put(337,-3){$\bigg]$}
%%%%%%%%%%%%%%%%%%%%%%%%%%%%%%%%%%%%%
\put(30,-53){$= - (\beta \hbar)^3 \frac{1}{8} \bigg \{ \frac{7}{180} \frac{1}{24}\bar{R} \Box \bar{R}
-\frac{1}{6} \frac{1}{144} \bar{R} \left( \bar{R}_{abcd} \right)^2 +\frac{7}{180} \bigg[\frac{1}{24}
D_{a} \bar{R}^{ab} D_{b} \bar{R} + \frac{1}{48} \left(D_a \bar{R} \right)^2 \bigg]$}
\put(30,-83){$+ \frac{7}{600}\bigg[\frac{1}{24}\bar{R} \Box \bar{R}+\frac{1}{12} \bar{R}D_{a}D_{b} \bar{R}^{ab}
-\frac{1}{27}\bar{R}\left(\bar{R}_{ab} \right)^2 -\frac{1}{18}\bar{R}\left(\bar{R}_{abcd} \right)^2 \bigg] 
\bigg \}$}
%%%%%%%%%%%%%%%%%%%%%%%%%%%%%%%%%%%%%
\end{picture}
\vspace{1.2in}

\item
%%\color{red} 
\beqr
I_5 &=& -\frac{1}{2}\left(\frac{\beta \hbar}{8} \right)^2 < \int_{-1}^{0}\int_{-1}^{0} \left(1-4\xi \right)^2 
 \bigg[ \left[D_{i_1}\bar{R}q^{i_1}\right](\sigma) \left[D_{i_2}\bar{R} q^{i_2} \right](\tau) \non \\
&+& \left[q^{i_1} q^{i_2}D_{i_1}D_{i_2}\bar{R}\right](\sigma) \bar{R} \bigg] \non \\
&+& 2\left(1-4\xi \right) \bar{R} 
\bigg[ \bar{g}^{ij} \partial_{i_1}
\bar{\Gamma}^{l}_{\hspace{0.05in}i k}\partial_{i_2}\bar{\Gamma}^{k}_{\hspace{0.05in}l j} q^{i_1} q^{i_2}\bigg](\tau)  
d\sigma d\tau >
\label{vertex5}
\feqr

\vspace{0.2in}

%%\color{blue}
\begin{picture}(0,0)
%%\color{green}
%%\graphpaper(-10,-30)(450,60)
%%\color{blue}
\thicklines
%%%%%%%%%%%%%%%%%%%%%%%%%%%%%%%%%%
\put(30,0){\circle*{3}}
\put(30,0){\line(1,0){20}}
\put(50,0){\circle*{3}}
%%%%%%%%%%%%%%%%%%%%%%%%%%%%%%%%%%%%%
\put(55,-3){$+$}
\put(67,-3){$\bigg[$}
%%%%%%%%%%%%%%%%%%%%%%%%%%%%%%%%%%%%%
\put(90,0){\circle{20}}
\put(100,0){\circle*{3}}
\put(105,-3){$\times$}
\put(120,0){\circle*{3}}
\put(127,-3){$\bigg]$}
\put(140,-3){$= \left(\beta \hbar\right)^3 \frac{1}{128}\bigg[\left(\frac{7}{15} \right)^2 \left(\frac{1}{12} 
D_{a}\bar{R}D^{a}\bar{R} +\frac{1}{6} \bar{R} \Box \bar{R}\right)
- \frac{14}{15}\frac{1}{36} \bar{R}(\bar{R}_{abcd})^2 \bigg]$}
\end{picture}

\vspace{0.5in}

\item 
%%\color{red}
\beqr
I_6 &=& \frac{1}{1296} \left(\frac{1}{\beta \hbar}\right)^3 < \int_{-1}^{0}\int_{-1}^{0}\int_{-1}^{0}
 \bigg[ \bar{R}_{i i_1 i_2 j} q^{i_1}q^{i_2} 
\left(\dot{q}^{i}\dot{q}^{j}+ b^{i}c^{j}+ a^{i}a^{j} \right)\bigg](\sigma) \non \\
&\times& \bigg[ \bar{R}_{k i_3 i_4 l} q^{i_3}q^{i_4} 
\left(\dot{q}^{k}\dot{q}^{l}+ b^{k}c^{l}+ a^{k}a^{l} \right)\bigg](\tau) \non \\
&\times& \bigg[ \bar{R}_{m i_5 i_6 n} q^{i_5}q^{i_6} 
\left(\dot{q}^{m}\dot{q}^{n}+ b^{m}c^{n}+ a^{m}a^{n} \right)\bigg](\rho) d\sigma d\tau d\rho >
\label{vertex6}
\feqr

\vspace{0.2in}

%%\color{blue}
\begin{picture}(0,0)
%%\color{green}
%%\graphpaper(-10,-170)(450,180)
%%\color{blue}
\thicklines
\put(27,-3){$\bigg[$}
\put(50,0){\circle{20}}
\put(60,0){\circle*{3}}
\put(70,0){\circle{20}}
\put(88,-3){$\bigg]^3$}
%%%%%%%%%%%%%%%%%%%%%%%%%%%%%%%%%%%%%%%
\put(95,-3){$+$}
%%%%%%%%%%%%%%%%%%%%%%%%%%%%%%%%%%%%%%%
\put(120,0){\circle{20}}
\put(130,0){\circle*{3}}
\put(140,0){\circle{20}}
\put(150,0){\circle*{3}}
\put(160,0){\circle{20}}
\put(170,0){\circle*{3}}
\put(180,0){\circle{20}}
%%%%%%%%%%%%%%%%%%%%%%%%%%%%%%%%%%%%%%
\put(195,-3){$+$}
\put(205,-3){$\bigg[$}
%%%%%%%%%%%%%%%%%%%%%%%%%%%%%%%%%%%%%5
\put(230,0){\circle{20}}
\put(220,0){\circle*{3}}
\put(240,0){\circle*{3}}
\qbezier(220,0)(230,8)(240,0)
\qbezier(220,0)(230,-8)(240,0)
%%%%%%%%%%%%%%%%%%%%%%%%%%%%%%%%%%%%
\put(245,-3){$\times$}
\put(270,0){\circle{20}}
\put(280,0){\circle*{3}}
\put(290,0){\circle{20}}
\put(307,-3){$\bigg]$}
%%%%%%%%%%%%%%%%%%%%%%%%%%%%%%%%%%%%%
\put(315,-3){$+$}
\put(340,0){\circle{20}}
\put(350,0){\circle*{3}}
\put(380,0){\circle{20}}
\put(380,10){\circle*{3}}
\put(380,-10){\circle*{3}}
\put(380,-10){\line(0,1){20}}
\qbezier(350,0)(365,10)(380,10)
\qbezier(350,0)(365,-10)(380,-10)
%%%%%%%%%%%%%%%%%%%%%%%%%%%%%%%%
\put(26,-53){$+$}
%%%%%%%%%%%%%%%%%%%%%%%%%%%%%%%%%
\put(38,-53){$\bigg[$}
\put(60,-50){\circle{20}}
\put(80,-50){\circle{20}}
\put(70,-50){\circle*{3}}
\put(95,-53){$\times$}
\put(120,-50){\circle{20}}
\put(140,-50){\circle{20}}
\put(160,-50){\circle{20}}
\put(130,-50){\circle*{3}}
\put(150,-50){\circle*{3}}
\put(177,-53){$\bigg]$}
%%%%%%%%%%%%%%%%%%%%%%%%%%%%%%%%%%%%%%%%
\put(185,-53){$+$}
\put(198,-53){$\bigg($}
%%%%%%%%%%%%%%%%%%%%%%%%%%%%%%%%%%%%%%%%
\put(230,-50){\circle{40}}
\put(210,-50){\circle*{3}}
\put(220,-50){\circle{20}}
\put(230,-50){\circle*{3}}
\put(240,-50){\circle{20}}
\put(250,-50){\circle*{3}}
%%%%%%%%%%%%%%%%%%%%%%%%%%%%%%%%%%%%%
\put(260,-53){$\equiv$}
%%%%%%%%%%%%%%%%%%%%%%%%%%%%%%%%%%%%%%
\put(300,-50){\circle{40}}
\put(280,-50){\line(1,1){20}}
\put(280,-50){\line(1,0){40}}
\put(320,-50){\line(-1,1){20}}
\put(300,-30){\circle*{3}}
\put(280,-50){\circle*{3}}
\put(320,-50){\circle*{3}}
%%%%%%%%%%%%%%%%%%%%%%%%%%%%%%%%%%%%%%
\put(325,-53){$\bigg)$}
\put(335,-53){$+$}
%%%%%%%%%%%%%%%%%%%%%%%%%%%%%%%%%%%%
\put(370,-50){\circle{20}}
\put(370,-35){\circle{10}}
\put(385,-50){\circle{10}}
\put(355,-50){\circle{10}}
\put(360,-50){\circle*{3}}
\put(370,-40){\circle*{3}}
\put(380,-50){\circle*{3}}
%%%%%%%%%%%%%%%%%%%%%%%%%%%%%%%%%%
\put(30,-93){$= \left(\beta \hbar \right)^3 \frac{1}{1296} \bigg[ \frac{1}{64} \bar{R}^3  
+ \frac{37}{360} \bar{R}^{ab} \bar{R}^{cd} \bar{R}_{acbd}
-\frac{3}{16} \left(\bar{R}_{abcd} \right)^2  \bar{R}
- \frac{40751}{10080} \bar{R}^{ab}\bar{R}^{cde}_{\hspace{0.2in} a} \bar{R}_{bcde}
$}
\put(30,-123){$ -\frac{1}{8} \bar{R} \left(\bar{R}_{ab} \right)^2
- \frac{20911}{10080} \bar{R}^{abcd} \bar{R}^{ef}_{\hspace{0.1in} ab} \bar{R}_{cdef} 
+\frac{409}{280} \bar{R}^{abcd} \bar{R}^{e \hspace{0.1in} f}_{\hspace{0.05in} a \hspace{0.1in} c}  \bar{R}_{bedf}
-\frac{137}{2016} \bar{R}^{ab} \bar{R}^{c}_{b} \bar{R}_{ca} \bigg] $}
%%%%%%%%%%%%%%%%%%%%%%%%%%%%%%%%%
\end{picture}

\vspace{1.8in}

\item 
%%\color{red}
\beqr
I_7 &=& -\frac{1}{576} \frac{1}{\beta \hbar} \left(1-4\xi\right) 
< \int_{-1}^{0}\int_{-1}^{0} \int_{-1}^{0} \bigg[ \bar{R}_{i i_1 i_2 j} q^{i_1}q^{i_2} 
\left(\dot{q}^{i}\dot{q}^{j}+ b^{i}c^{j}+ a^{i}a^{j} \right)\bigg](\sigma) \non \\
&\times& \bigg[ \bar{R}_{k i_3 i_4 l} q^{i_3}q^{i_4} 
\left(\dot{q}^{k}\dot{q}^{l}+ b^{k}c^{l}+ a^{k}a^{l} \right)\bigg](\tau) \bar{R}
d\sigma d\tau d\rho >
\label{vertex7}
\feqr

\vspace{0.2in}

%%\color{blue}
\begin{picture}(0,0)
%%\color{green}
%%\graphpaper(-10,-40)(450,70)
%%\color{blue}
\thicklines
\put(28,-3){$\bigg[$}
\put(38,-3){$\bigg($}
\put(60,0){\circle{20}}
\put(70,0){\circle*{3}}
\put(80,0){\circle{20}}
\put(97,-3){$\bigg)^2$}
\put(105,-3){$\times$}
\put(120,0){\circle*{4}}
\put(127,-3){$\bigg]$}
%%%%%%%%%%%%%%%%%%%%%%%%%%%%%%%%%%%%%%%
\put(135,-3){$+$}
\put(147,-3){$\bigg[$}
%%%%%%%%%%%%%%%%%%%%%%%%%%%%%%%%%%%%%%%
\put(170,0){\circle{20}}
\put(180,0){\circle*{3}}
\put(190,0){\circle{20}}
\put(200,0){\circle*{3}}
\put(210,0){\circle{20}}
%%%%%%%%%%%%%%%%%%%%%%%%%%%%%%%%%%%%%%
\put(225,-3){$\times$}
\put(240,0){\circle*{4}}
\put(247,-3){$\bigg]$}
%%%%%%%%%%%%%%%%%%%%%%%%%%%%%%%%%%%%%5
\put(255,-3){$+$}
\put(267,-3){$\bigg[$}
\put(290,0){\circle{20}}
\put(280,0){\circle*{3}}
\put(300,0){\circle*{3}}
\qbezier(280,0)(290,8)(300,0)
\qbezier(280,0)(290,-8)(300,0)
\put(305,-3){$\times$}
\put(320,0){\circle*{4}}
\put(327,-3){$\bigg]$}
\put(30,-33){$= - \left(\beta \hbar \right)^3 \frac{1}{576} \frac{7}{15} \bar{R} 
\bigg[\frac{1}{16} \bar{R}^2 - \frac{1}{6} \left(R_{ab} \right)^2 -\frac{1}{4}\left(R_{abcd} \right)^2 \bigg]$}
\end{picture}

\vspace{0.5in}

\item 
%%\color{red}
\beqr
I_8 &=& \frac{1}{768}  \beta \hbar \left(1-4\xi \right)^2 < \int_{-1}^{0}\int_{-1}^{0}
\int_{-1}^{0} \bigg[ \bar{R}_{i i_1 i_2 j} q^{i_1}q^{i_2} 
\left(\dot{q}^{i}\dot{q}^{j}+ b^{i}c^{j}+ a^{i}a^{j} \right)\bigg](\sigma) \non \\
&\times& \bar{R} \bar{R} d\sigma d\tau d\rho >
\label{vertex8}
\feqr

\vspace{0.2in}
%%\color{blue}
\begin{picture}(0,0)
%%\color{green}
%%\graphpaper(-10,-30)(450,60)
%%\color{blue}
\thicklines
\put(28,-3){$\bigg[$}
\put(40,0){\circle*{4}}
\put(47,-3){$\bigg]^2$}
\put(55,-3){$\times$}
\put(80,0){\circle{20}}
\put(90,0){\circle*{3}}
\put(100,0){\circle{20}}
\put(120,-3){$=\left(\beta \hbar\right)^3 \frac{1}{768} \left(\frac{7}{15}\right)^2 \frac{1}{4}\bar{R}^3$}
\end{picture}

\vspace{0.5in}

\item 
%%\color{red}
\beqr
I_9 &=& \frac{1}{3!} \left(\frac{- \beta \hbar}{8}\right)^3  \left(1-4\xi \right)^3 
< \int_{-1}^{0}\int_{-1}^{0}\int_{-1}^{0} \bar{R} \bar{R} \bar{R} d\sigma d\tau d\rho >
\label{vertex9}
\feqr

\vspace{0.2in}
%%\color{blue}
\begin{picture}(0,0)
%%\color{green}
%%\graphpaper(-10,-30)(450,60)
%%\color{blue}
\put(28,-3){$\bigg[$}
\put(40,0){\circle*{4}}
\put(47,-3){$\bigg]^3$}
\put(60,-3){$=-\frac{1}{3072} \left(\frac{7}{15}\right)^3 \left(\beta \hbar\right)^3 \bar{R}^3$}
\end{picture}

\vspace{0.5in}
%%\color{black}
 
Some comments are in order

\begin{itemize}

\item There are Feynman diagrams that satisfy the factorization property according to which a diagram breaks down 
into simpler subdiagrams. The four dimensional building block diagrams are depicted with their corresponding 
contributions in what follows:  

\vspace{1.5in}

%%\color{blue}
\begin{picture}(0,0)
%%\color{green}
%%\graphpaper(-10,-170)(450,180)
%%\color{blue}
\thicklines
%%%%%%%%%%%%%%%%%%%%%%%%%%%%%%%%%%%%%%%%%%%
\put(30,0){\circle{20}}
\put(40,0){\circle*{3}}
\put(50,0){\circle{20}}
\put(90,-3){$=\frac{1}{4} \bar{R}^2 $} 
%%%%%%%%%%%%%%%%%%%%%%%%%%%%%%%%%%%%%%%%%%%%%%
\put(30,-30){\circle{20}}
\put(40,-30){\circle*{3}}
\put(50,-30){\circle{20}}
\put(60,-30){\circle*{3}}
\put(70,-30){\circle{20}}
\put(90,-33){$=- \frac{1}{6} \left(R_{ab} \right)^2$}
%%%%%%%%%%%%%%%%%%%%%%%%%%%%%%%%%%%%%%%%%%%%%%%%
\put(40,-60){\circle{20}}
\put(30,-60){\circle*{3}}
\put(50,-60){\circle*{3}}
\qbezier(30,-60)(40,-68)(50,-60)
\qbezier(30,-60)(40,-52)(50,-60)
\put(90,-63){$=-\frac{1}{4}\left(R_{abcd} \right)^2$}
%%%%%%%%%%%%%%%%%%%%%%%%%%%%%%%%%%%%%%%%%%%%%%
\put(38,-106){\usebox{\ninety}}
\put(35,-112){\usebox{\twotwofive}}
\put(41,-113){\usebox{\thonefive}}
\put(40,-110){\circle*{3}}
\put(90,-113){$= \alpha \left( \frac{1}{16} \bar{R} \left(\bar{R}_{abcd} \right)^2 
+\frac{1}{24} \bar{R} \left(\bar{R}_{ab} \right)^2 \right) + \beta \frac{1}{12} \Box \bar{R} $}
\end{picture}

\vspace{2.5in}
%%\color{black}

\item Let us examine now the contribution of $\bar{R}^{\frac{n}{2}}$ terms. In n=2 dimensions the trace anomaly is:
\beqr
\alpha_2=-\frac{1}{24 \pi} \bar{R}.
\label{trace2}
\feqr

\noi The situation changes in four dimensions in which the $\bar{R}^2$ terms, steming from the interactions:
\beqr
I_{\bar{R}^2} &=& -\frac{1}{144} \bar{R}_{i i_1 i_2 j}\bar{R}
<\int_{-1}^{0} q^{i_1} q^{i_2} \left(\dot{q}^{i}\dot{q}^{j}+ b^{i}c^{j}+ a^{i}a^{j} \right) d\sigma> \non \\
&+& \frac{1}{72 \left(\beta \hbar \right)^2}\bar{R}_{k i_1 i_2 l}\bar{R}_{m i_3 i_4 n}
<\int_{-1}^{0} \int_{-1}^{0}
\bigg[ q^{i_1} q^{i_2} \left(\dot{q}^{k}\dot{q}^{l}+ b^{k}c^{l}+ a^{k}a^{l} \right)\bigg](\sigma) 
\non \\
&\times&
\bigg[ q^{i_3} q^{i_4} \left(\dot{q}^{m}\dot{q}^{n}+ b^{m}c^{n}+ a^{m}a^{n} \right)\bigg](\tau)
d\sigma d\tau> \non \\
&+& \frac{\left(\beta \hbar \right)^2}{1152}\bar{R}^2
\label{dim4R2}
\feqr

\noi and represented by the following diagrams:

\vspace{0.2in}

%%\color{blue}
\begin{picture}(0,0)
%%\color{green}
%%\graphpaper(-10,-70)(450,80)
%%\color{blue}
\thicklines
%%%%%%%%%%%%%%%%%%%%%%%%%%%%%%%%%%%%%%%%%%%
\put(30,0){\circle{20}}
\put(40,0){\circle*{3}}
\put(50,0){\circle{20}}
\put(90,-3){$=-\left(-\beta \hbar \right)^2 \left(\frac{1}{24}\right)^2 \bar{R}^2$} 
%%%%%%%%%%%%%%%%%%%%%%%%%%%%%%%%%%%%%%%%%%%%%%
\put(30,-30){\circle{20}}
\put(40,-30){\circle*{3}}
\put(50,-30){\circle{20}}
\put(60,-30){\circle*{3}}
\put(70,-30){\circle{20}}
\put(90,-33){$=\left(-\beta \hbar \right)^2 \left[ \frac{1}{2} \left(\frac{1}{24}\right)^2 \bar{R}^2
-\frac{1}{432}\left(\bar{R}_{ab}\right)^2 \right]$}
%%%%%%%%%%%%%%%%%%%%%%%%%%%%%%%%%%%%%%%%%%%%%%%%
\put(40,-60){\circle*{3}}
\put(90,-63){$=\left(-\beta \hbar \right)^2\frac{1}{2}\left(\frac{1}{24} \right)^2 \bar{R}^2$}
\end{picture}

\vspace{1.5in}
%%\color{black}

produce a vanishing result.

The same observation holds for the $\bar{R}$ terms created by the interaction:
\beqr
I_{\bar{R}} &=& \frac{1}{\beta \hbar} \frac{1}{6}\bar{R}_{i i_1 i_2 j}
<\int_{-1}^{0} q^{i_1} q^{i_2} \left(\dot{q}^{i}\dot{q}^{j}+ b^{i}c^{j}+ a^{i}a^{j} \right) d\sigma> \non \\
&-& \beta \hbar \frac{1}{24}\bar{R}
\label{dim4R}
\feqr

\noi and associated with the Feynman diagrams

\vspace{0.5in}
%%\color{blue}
\begin{picture}(0,0)
%%\color{green}
%%\graphpaper(-10,-30)(450,60)
%%\color{blue}
\thicklines
%%%%%%%%%%%%%%%%%%%%%%%%%%%%%%%%%%%%%%%%%%%%
\put(30,0){\circle{20}}
\put(40,0){\circle*{3}}
\put(50,0){\circle{20}}
%%%%%%%%%%%%%%%%%%%%%%%%%%%%%%
\put(65,-3){$+$}
\put(80,0){\circle*{3}}
\put(90,-3){$=\beta \hbar \bar{R} \left[\frac{1}{24}- \frac{1}{24}\right]$}
\end{picture}

\vspace{0.5in}
%%\color{black}

One making use of the identities \textit{(9)-(24)} of the appendix and the 17 linearly independent terms listed below:
\beqr
&& R^3, \quad R\left(R_{ab} \right)^2, \quad R\left(R_{abcd} \right)^2, \quad R_{ab}R^{b}_{c}R^{ac}, 
\quad R^{ab}R^{cd}R_{acbd} \non \\
&& R_{ab}R^{acde}R^{b}_{\hspace{0.05in}cde}, \quad R_{abcd}R^{abef}R^{cd}_{\hspace{0.1in}ef}, 
\quad R^{abcd} R^{e \hspace{0.1in} f}_{\hspace{0.05in} a \hspace{0.1in} c} R_{bedf}, \quad R\Box R \non \\
&& R_{ab} \Box R^{ab}, \quad R_{abcd}\Box R^{abcd}, \quad \left(D_{a}R_{bc} \right)^2, 
\quad R^{ab}D_{b}D_{c}R^{c}_{a}  \non \\
&& D^{a}R^{bc}D_{b}R_{ac}, \quad \left(D^{a}R^{bcde}\right)^2, \quad \Box^2 R, \quad \left(D_{a}R \right)^2 
\label{linearin}
\feqr
\end{itemize}
\end{enumerate}

\noi can deduce for the integrated trace anomaly:
\beqr
An_{W}(s=0, n) =\lim_{\beta \rightarrow 0}\left( \frac{1}{(2\pi \beta \hbar)^{3}}
\int dx^{i}_{0}\prod_{i=1}^{n} \sqrt{g(x^{i}_{0})} f(x_{0}^{i}) I(x^{i}_{0}) \right)
\label{finres}
\feqr

\noi where
\beqr
\frac{I(x^{i}_{0})}{(\beta \hbar)^{3}} &=& -\frac{1}{1296000}\bar{R}^3 +\frac{7}{129600}\bar{R}\left(\bar{R}_{ab} \right)^2 
+\frac{1}{43200}\bar{R}\left(\bar{R}_{abcd} \right)^2 
-\frac{5293}{13063680}\bar{R}_{ab}\bar{R}^{b}_{c}\bar{R}^{ac} \non \\
&+& \frac{9287}{16329600}\bar{R}^{ab}\bar{R}^{cd}\bar{R}_{acbd} 
- \frac{159421}{130636800}\bar{R}_{ab}\bar{R}^{acde}\bar{R}^{b}_{\hspace{0.05in}cde}
-\frac{18413}{26127360}\bar{R}_{abcd}\bar{R}^{abef}\bar{R}^{cd}_{\hspace{0.1in}ef} \non \\
&+&\frac{661}{362880}
\bar{R}^{abcd} \bar{R}^{e \hspace{0.1in} f}_{\hspace{0.05in} a \hspace{0.1in} c} \bar{R}_{bedf}
+\frac{1}{21600}\bar{R}\Box \bar{R} 
-\frac{3}{5600} \bar{R}_{ab} \Box \bar{R}^{ab} \non \\ 
&-& \frac{191}{483840}\bar{R}_{abcd}\Box \bar{R}^{abcd} +
\frac{7}{8640}D^{a}\bar{R}^{bc}D_{b}\bar{R}_{ac}-\frac{1}{20160}\left(D^{a}\bar{R}^{bcde}\right)^2 \non\\
&-&\frac{17}{100800}\Box^2 \bar{R} 
- \frac{1}{5040} \bar{R}^{ab}D_{b}D_{c}\bar{R}^{c}_{a} 
-\frac{127}{120960}\left(D_{a}\bar{R}_{bc} \right)^2 -\frac{67}{604880}\left(D_{a}\bar{R} \right)^2 .
\label{Iexp}
\feqr

%%%%%%%%%%%%%%%%%%%%%%%%%%%%%%%%%%%%%%%%%%%%%%%%%%%%%%%%%%%%%%%%%%%%%%%%%%%%%%%%%%%%%%%%%%%%%%%%%%%%%%%%%%%%%%%%%%%%%%%%%%%

                                             %%%%%%%%%%%%%%%%%%%%%%
                                             % COMPUTER ALGEBRA   %
                                             % PROGRAM            %
                                             %%%%%%%%%%%%%%%%%%%%%%

%%%%%%%%%%%%%%%%%%%%%%%%%%%%%%%%%%%%%%%%%%%%%%%%%%%%%%%%%%%%%%%%%%%%%%%%%%%%%%%%%%%%%%%%%%%%%%%%%%%%%%%%%%%%%%%%%%%%%%%%%%

\section{The computer algebra program}

Vertices $I_1$ to $I_9$ were calculated using the 
Riegeom package \cite{Riegeom},
%%% ref at the end %%%
which is a Maple package for manipulating generic symbolic tensor
expressions in the context of Riemannian geometry.
In this section we show the process of calculating
the vertex $I_7$.
The lines beginning with ``$>$'' in typewrite font are 
the input in a Maple worksheet. A colon at the end
of the command hides the result.
We start loading the package \\
\noindent
$>$ \texttt{with(Riegeom);}\\
{\footnotesize defining Christoffel(Gamma), Riemann(R), Weyl(C), Ricci(R), 
LeviCivita(eta),\\
 TraceFreeRicci(S) for Dimension = 4, CoordinateName =
X, MetricName = g}
\begin{eqnarray*}
\lefteqn{[\mathit{absorbg}, \,\mathit{changedumind}, \,\mathit{
cleartensor}, \,\mathit{codiff}, \,\mathit{coordinate}, \,
\mathit{definetensor}, \,\mathit{dimension}, } \\
 & & \mathit{expandcodiff}, \,\mathit{lptensor}, \,\mathit{
ltensor}, \,\mathit{metric}, \,\mathit{normalform}, \,\mathit{off
}, \,\mathit{on}, \,\mathit{printtensor}, \,\mathit{replace},  \\
 & & \mathit{simpLC}, \,\mathit{simptensor}, \,\mathit{sreplace}
, \,\mathit{switches}, \,\mathit{symmetrize}, \,\mathit{symmetry}
, \,\mathit{tdiff}]
\end{eqnarray*}
Next command reads file ``Vertex.mpl'' that contains 
procedures written specifically for the calculation 
of vertices $I_1$ to $I_9$ using Maple programming
language and Riegeom commands. \\
$>$ \texttt{read(`Vertex.mpl`):} \\
We setup spacetime dimension using Riegeom 
command \texttt{dimension}. \\
\noindent
$>$ \texttt{dimension(6):} \\
Next command enters the tensor coefficient of vertex $I_7$. \\
$>$ \texttt{tensor$\_$coeff :=
printtensor(R[-mu1,-i1,-i2,-nu1]*R[-mu2,-i3,-i4,-nu2]);}{%
}
\[
\mathit{tensor\_coeff} := R\,{_{\mu 1}}\,{_{\mathit{i1}}}\,{
_{\mathit{i2}}}\,{_{\nu 1}}\,R\,{_{\mu 2}}\,{_{\mathit{i3
}}}\,{_{\mathit{i4}}}\,{_{\nu 2}}
\]
\texttt{Printtensor} is the Riegeom interface command. Indices with
minus sign are covariant and with plus sign are contravariant.
In the next command we enter the field terms. 
We use a Maple list (which preserves order)
instead of an expression written in terms of the commuting product operator.
In this form we have full control over the order of the fields. \\
$>$ \texttt{L := printtensor([[q[i1],q[i2]](sigma),[[qdot[mu1],qdot[nu1]],}\\
$>$ \,\,\texttt{[b[mu1],c[nu1]],
[a[mu1],a[nu1]]](sigma),[q[i3],q[i4]](tau),}\\
$>$ \,\,\texttt{[[qdot[mu2],qdot[nu2]],
[b[mu2],c[nu2]],
[a[mu2],a[nu2]]](tau)]);}
\begin{eqnarray*}
\lefteqn{L := [[q\ ^{\mathit{i1}}(\sigma ), q
\ ^{\mathit{i2}}(\sigma )], } \\
 & & [[\mathit{qdot}\ ^{\mu 1}(\sigma ), 
\mathit{qdot}\ ^{\nu 1}(\sigma )], [b\ ^{\mu 1
}(\sigma ), c\ ^{\nu 1}(\sigma )], 
[a\ ^{\mu 1}(\sigma ), a\ ^{\nu 1}\mathrm{
\ }(\sigma )]],  \\
 & & [q\ ^{\mathit{i3}}(\tau ), q\ ^{\mathit{
i4}}(\tau )],  \\
 & & [[\mathit{qdot}\ ^{\mu 2}(\tau ), \mathit{
qdot}\ ^{\nu 2}(\tau )], [b\ ^{\mu 2}
(\tau ), c\ ^{\nu 2}(\tau )], [a
\ ^{\mu 2}(\tau ), a\ ^{\nu 2}(\tau
 )]]]
\end{eqnarray*}
Next command finds all independent \footnote{The number of independent index configurations is given by 
$\frac{n!}{\left(\frac{n}{2} \right)! \, \, 2^{\frac{n}{2}}}$ where $\left(\frac{n}{2} \right)!$ represents 
the possible permutations of index pairs and $2^{\frac{n}{2}}$ the permutation of indices within the pairs.}
index configurations that contribute
to the final result. \\
$>$ \texttt{Lic :=
ind$\_$config([i1,i2,mu1,nu1,i3,i4,mu2,nu2]):} \\
$>$ \texttt{N := nops(Lic);}
\[
N := 105
\]
In the case of 8 field indices, there are 105 independent index configurations.
We show the first 3 ones.

$>$ \texttt{for i to 3 do evaln(Lic[i])=Lic[i] od;}
$$
{\mathit{Lic}_{1}}=[\mathit{i1}, \,\mathit{i2}, \,\mu 1, \,
\nu 1, \,\mathit{i3}, \,\mathit{i4}, \,\mu 2, \,\nu 2]
$$
$$
{\mathit{Lic}_{2}}=[\mathit{i1}, \,\mathit{i2}, \,\mu 1, \,
\nu 1, \,\mathit{i3}, \,\mu 2, \,\mathit{i4}, \,\nu 2]
$$
$$
{\mathit{Lic}_{3}}=[\mathit{i1}, \,\mathit{i2}, \,\mu 1, \,
\nu 1, \,\mathit{i3}, \,\nu 2, \,\mu 2, \,\mathit{i4}]
$$
Next command is a loop over the 105 index configurations. The
results are stored in a table of results called \texttt{res}. \\
$>$ \texttt{for i to N do } \\
$>$ \texttt{\,\,\,   expr := WickContractions(L, Lic[i]):} \\
$>$ \texttt{\,\,\,   res[i] := Vertex(tensor$\_$coeff*expr):} \\
$>$ \texttt{od:} \\
Command \texttt{WickContractions} performs the Wick contractions
given by eq. (27) to (32). We show an example with the
fourth index configuration.\\
$>$ \texttt{expr :=
factor(WickContractions(L, Lic[4]));}{%
}
\[
\mathit{expr} := \beta ^{4}\,\mathit{hbar}^{4}\,
g\,^{\mathit{i1}}\,^{\mathit{i2}}\,g\,^{
\mu 2}\,^{\nu 2}\,g\,^{\mu 1}\,^{\mathit{i3}}\,g\,^{\nu 1
}\,^{\mathit{i4}}\,\sigma \,(1 + \sigma )\,(
\tau  + \mathrm{Heaviside}(\sigma  - \tau ))^{2}
\]
Command \texttt{Vertex} simply multiplies the result of Wick
contractions to the tensor coefficient, simplifies the
tensor expression, and isolates the terms to be integrated,
since we noticed that, for the most complicate vertices, Maple spends
more time integrating Dirac delta and step functions than 
simplifying the tensorial terms.   \\
$>$ \texttt{Vertex(tensor$\_$coeff*expr);}
\[
[\beta ^{4}\,\mathit{hbar}^{4}\,R\,^{\nu 2}\,^{\nu 1}\,R\,{
_{\nu 2}}\,{_{\nu 1}}, \,\sigma \,(1 + \sigma )\,(\tau  + 
\mathrm{Heaviside}(\sigma  - \tau ))^{2}]
\]
Next command adds the results of all index configurations after
integrating over variables $\sigma$ and $\tau$. Functions 
\texttt{int$\_$tau} and \texttt{int$\_$sigma} replace the usual
Maple integrator, since Maple \texttt{int}
command fails to return the correct value for complicated vertices.  \\
$>$ \texttt{final$\_$value :=
simptensor(add(res[i][1]*}\\
$>$ \texttt{int$\_$tau(int$\_$sigma(res[i][2])),i=1..N));}{%
}
\begin{eqnarray*}
\mathit{final\_value} :=   \frac {1}{16} \,R^{2}\,\beta ^{4}
\,\mathit{hbar}^{4} -
 \frac {1}{6} \,\beta ^{4}\,\mathit{hbar}^{4}\,R\,
^{\nu 1}\,^{\mu 2}\,R\,{_{\nu 1}}\,{_{\mu 2}} -\\
\frac {1}{4} \,
\beta^{4}\,\mathit{hbar}^{4}
\,R\,^{\nu 2}\,^{\mathit{i1}}\,^{\mathit{\_b0}}\,
^{\mathit{\_c0}}\,R\,{_{\nu 2}}\,{_{\mathit{i1}}}\,
{_{\mathit{\_b0}}}\,{_{\mathit{\_c0}}} \mbox{\hspace{1.3cm}} 
\end{eqnarray*}
Collecting $\hbar^4\beta^4$, we obtain the final form. \\
$>$ \texttt{collect(final$\_$value, [beta,hbar]);}
\[
(  {\displaystyle \frac {1}{16}} \,R^{2}
- {\displaystyle \frac {1}{6}} \,R\,^{\nu 1}\,^{\mu 2}\,R\,{
_{\nu 1}}\,{_{\mu 2}}
- {\displaystyle \frac {1}{4}} \,R\,^{\nu 2}\,^{\mathit{i1}
}\,^{\mathit{\_b0}}\,^{\mathit{\_c0}}\,R\,{_{\nu 2}}\,{_{
\mathit{i1}}}\,{_{\mathit{\_b0}}}\,{_{\mathit{\_c0}}}  )
\,\mathit{hbar}^{4}\,\beta ^{4}
\]
For vertices $I_2$ and $I_6$, the method described above
has some extra complications. Vertex $I_6$ has a large 
number of index configurations ($N := 10395$) and vertex  $I_2$ uses
huge side identities obtained in Riemann normal 
coordinates as described in appendix $A.3$.

This computer algorithm can easily be extended to higher dimensions to predict 
the trace anomalies in 8 and 10 dimensions. 
Of course there are strong limitations implied by the exponential growth of the problem which can be relaxed by 
increasing the available computing power. The tensorial upper bound of the program is reached by the product of 
eight Riemann curvature tensors but such a case is beyond the scope of the present work.

%%%%%%%%%%%%%%%%%%%%%%%%%%%%%%%%%%%%%%%%%%%%%%%%%%%%%%%%%%%%%%%%%%%%%%%%%%%%%%%%%%%%%%%%%%%%%%%%%%%%%%%%%%%%%%%%%

%%%%%%%%%%%%%%%%%%%%%%%%%%%%%%%%%%%%%%%%%%%%%%%%%%%%%%%%%%%%%%%%%%%%%%%%%%%%%%%%%%%%%%%%%%%%%%%%%%%%%%%%%%%%%%%%%%%%%%%%%%%

						%%%%%%%%%%%%%%%%%%
						%   CONCLUSIONS  %
						%%%%%%%%%%%%%%%%%%

\section{Conclusions}

In this article we calculate the integrated trace anomaly for a real spin-0 
scalar field living on a curved six dimensional manifold. This is achieved by relying on a recursive
computation of the metric tensor components in Riemann normal coordinates. One  
can use the general formulae \rf{genex} and \rf{coeexp} to reach the desired order of metric expansion 
induced  by the dimensionality of the manifold. Adopting the path 
integral formalism of quantun mechanical non-linear sigma models, we evaluate all the vertices and the corresponding 
Feynman diagrams that contribute to the present anomaly. A computer based program is used to perform the otherwise 
inevitable task of integration over distributions and contractions of the various tensors involved. The final 
result derived in this process involves 17 scalar monomials consisting of covariant derivatives and/or Riemann tensors.

%%%%%%%%%%%%%%%%%%%%%%%%%%%%%%%%%%%%%%%%%%%%%%%%%%%%%%%%%%%%%%%%%%%%%%%%%%%%%%%%%%%%%%%%%%%%%%%%%%%%%%%%%%%%%%%%%%%%%%%%%%%%

						%%%%%%%%%%%%%%%%%%
						%   Appendix  %
						%%%%%%%%%%%%%%%%%%

\vspace{0.5 cm}
\appendix
\section*{Appendix}
\setcounter{section} {1}
\setcounter{equation} {0}
\indent

\subsection{Useful identities}

We consider a Riemannian (or pseudo-Riemannian) manifol equipped with its standard (metric compatible, torsion free) 
connection. The Riemann curvature is defined as:
\beqr
R^{a}_{\hspace{0.1in} bcd} = \partial_{c} \Gamma^{a}_{\hspace{0.05in}bd}
+\Gamma^{e}_{\hspace{0.05in}bd} \Gamma^{a}_{\hspace{0.05in}e c} - c\leftrightarrow d
\label{riemcur}
\feqr

\noi possessing the familiar symmetries:
\begin{itemize}
\item antisymmety
\beqr
R_{abcd}=-R_{bacd}=-R_{abdc}
\label{anti1}
\feqr

\item pair symmetry
\beqr
R_{abcd}=R_{badc}
\label{pais1}
\feqr

\item cyclic symmetry
\beqr
R_{abcd}+R_{adbc}+R_{acdb}=0
\label{cyclics}
\feqr

\item Bianchi symmetry and the related identities
\beqr
R_{abcd; e}+R_{abec; d}+R_{abde; c} &=& 0 \\
R_{bc;a}-R_{ab;c} + R^{d}_{\hspace{0.1in} bca;d}&=& 0 \\
\left(R^{a}_{b}-\frac{1}{2}\delta^{a}_{b}R \right)_{;a} &=& 0
\label{bianchis}
\feqr

\end{itemize}

The Ricci curvature and scalar are:
\beqr
R_{ab}=R^{q}_{\hspace{0.1in}aqb}; \quad R=R^{q}_{\hspace{0.1in}q} 
\label{riccics}
\feqr

\noi In this paper the following identities have been exploited to simplify our expressions:
\beqr
R_{abcd}R^{acbd} &=& \frac{1}{2} R^{2}_{abcd} \\
R_{abcd}R^{ecbd} &=& \frac{1}{2} R_{abcd}R^{ebcd} \\
R_{a(bc)d} R^{e(af)b} R^{\hspace{0.05in} (d \hspace{0.1in} c}_{f \hspace{0.05in} e)} &=& \frac{1}{8}
R_{abcd} \left[R^{e \hspace{0.1in} c}_{\hspace{0.05in} af}R^{b \hspace{0.1in} f}_{\hspace{0.05in} ed}
+\frac{7}{2}R^{ab ef}R_{ef}^{\hspace{0.1in} cd} \right] \\
\Box D_{a}D_{b}R^{ab} &=& \frac{1}{2} \Box^{2} R \\
D_{a}D_{b}\Box R^{ab}&=& \frac{1}{2} \Box^{2} R - \frac{1}{2}\left(D_{a}R \right)^2 - 2 R_{ab}D^{a}D^{b} R \non \\
&+& 2 R_{ab}R^{b}_{c}R^{ac}-2R^{ab}R^{cd}R_{acbd}-4 D_{a}R_{bc}D^{b}R^{ac} \non \\
&+& 3 \left(D_{a}R_{bc} \right)^2 + \frac{1}{2}R_{abcd}\Box R^{abcd}
-2R_{abcd}R^{e \hspace{0.2in}c}_{\hspace{0.1in}af}R^{be}_{\hspace{0.1in}df} \non \\
&+& 2 R^{ab}R^{cde}_{\hspace{0.2in}a}R_{bcde}-\frac{1}{2}R^{abcd}R_{abef}R_{cd}^{\hspace{0.1in}ef}
+ R_{ab}\Box R^{ab}\\
D_{a}D_{b}D^{a}D_{c}R^{bc} &=& \frac{1}{2} \Box^{2} R -\frac{1}{4}\left(D_{a}R \right)^2
-\frac{1}{2}R_{ab}D^{a}D^{b}R \\
D_{a}D_{b}D_{c}D^{b}R^{ac} &=& \frac{1}{2} \Box^{2} R -\frac{1}{2}\left(D_{a}R \right)^2 
- 2R_{ab}D^{a}D^{b}R \non \\
&+& R^{abcd}D_{d}D_{b}R_{ac}+2R_{ab}R^{b}_{c}R^{ac}-2R^{ab}R^{cd}R_{acbd} \non \\
&-& 3D_{a}R_{bc}D^{b}R^{ac}+R_{ab}\Box R^{ab}+2\left(D_{a}R_{bc} \right)^2 \\
D_{a}D_{b}D_{c}D^{a}R^{bc} &=& \frac{1}{2} \Box^{2} R -\frac{1}{2}\left(D_{a}R \right)^2 
- 2R_{ab}D^{a}D^{b}R \non \\
&+& 2R_{ab}R^{b}_{c}R^{ac}-2R^{ab}R^{cd}R_{acbd} - 3D_{a}R_{bc}D^{b}R^{ac} \non \\
&+&R_{ab}\Box R^{ab}+2\left(D_{a}R_{bc} \right)^2+\frac{1}{4}R_{abcd}\Box R^{abcd} \non \\
&-&R^{abcd}R^{e \hspace{0.2in}c}_{\hspace{0.1in}af}R^{be}_{\hspace{0.1in}df}
+R^{ab} R^{cde}_{\hspace{0.2in}a}R_{bcde}-\frac{1}{4}R^{abcd}R_{abef}R_{cd}^{\hspace{0.1in}ef} \non \\
D^{a}R^{bc}D_{d}R^{\hspace{0.1in} d}_{ab \hspace{0.1in} c} &=& \left(D_{a}R_{bc} \right)^2 -D_{a}R_{bc}D^{b}R^{ac} \\
R^{abcd}D_{d}D_{b}R_{ac} &=& \frac{1}{4}R^{abcd}\Box R_{abcd}+\frac{1}{2}R_{ab}R^{acde}R^{b}_{\hspace{0.1in}cde}-
R_{abcd}R^{e \hspace{0.2in}c}_{\hspace{0.1in}af}R^{bf}_{\hspace{0.1in}de} \\
R^{ab}D^{c}D_{d}R^{\hspace{0.1in} d}_{ca \hspace{0.1in} b} &=& R^{ab}\Box R_{ab}-\frac{1}{2}R^{ab}D_{a}D_{b}R
+R^{ab}R^{c}_{a}R_{cb}-R^{ab}R^{cd}R_{acbd}\\
R^{abcd}D_{d}D_{e}R^{e}_{\hspace{0.1in}abc} &=& R^{abcd}D_{d}D_{b}R_{ac}\\
D_{a}R^{abcd}D_{e}R^{e}_{\hspace{0.1in}bcd} &=& 2 \left[\left(D_{a}R_{bc} \right)^2 - D_{a}R_{bc}D^{b}R^{ac}\right] \\
R^{abcd}D_{e}D_{d}R^{e}_{\hspace{0.1in}abc} &=& \frac{1}{4} R^{abcd}\Box R_{abcd} \\
D^{a}R^{ebcd}D_{e}R_{abcd} &=& \frac{1}{2} \left(D_{a}R_{bcde} \right)^2 \\
D_{a}R D^{a}R &=& \frac{1}{2}\Box R^2 - R\Box R \\
R^{ab}D_{b}D_{a}R &=& 2R^{ab}D^{c}D_{b}R_{ac}-2R^{ab}R^{cd}R_{acbd}+2R^{ab}R^{c}_{a}R_{bc}
%\\
%R_{abcd}\Box R^{abcd} &=& 4R^{abcd}D_{a}D_{c}R_{bd}-2R_{ab}R^{acde}R^{b}_{\hspace{0.1in}cde}+
%4R_{abcd}R^{e \hspace{0.2in}c}_{\hspace{0.1in}af}R^{bf}_{\hspace{0.1in}de}
\label{ident1}
\feqr

\noi \textit{Proof of (9)} By cyclic symmetry,
\beqr
R_{abcd}R^{acbd}=-R_{abcd}\left( R^{adcb}+R^{abdc} \right) =-R_{abcd}R^{adcb}+R^{2}_{abcd}.
\label{prcy}
\feqr

\noi Rename the indices in the first term:
\beqr
R_{abcd}R^{adcb}=R_{abdc}R^{acdb}=R_{abcd}R^{acbd}
\label{rena}
\feqr

\noi Solve for the desired object. In the same vein we can prove (10) as well as $D_{a}R^{abcd}D^{e}R_{ecbd}=
\frac{1}{2}\left(D_{a}R^{abcd} \right)^2$ and $R^{abcd} \Box R_{acbd}=\frac{1}{2}R^{abcd} \Box R_{abcd}$.

\textit{Proof of (11)} Expanding out the products, the resulting sum can be expressed in terms of 
the invariants:
\beqr
L_1 &=& R_{abcd}R^{abef}R^{c \hspace{0.1in} d}_{\hspace{0.05in} e  \hspace{0.1in} f} \non \\
L_2 &=& R_{abcd}R^{ab ef}R_{ef}^{\hspace{0.1in} cd} \non \\
L_3 &=& R_{abcd}R^{e \hspace{0.1in} c}_{\hspace{0.05in} fa}R^{f \hspace{0.1in} b}_{\hspace{0.05in} de} \non \\
L_4 &=& R_{abcd}R^{e \hspace{0.1in} c}_{\hspace{0.05in} af}R^{b \hspace{0.1in} f}_{\hspace{0.05in} ed}
\label{equite}  
\feqr

\noi and with the assistance of the identities:
\beqr
L_1 &=& \frac{1}{2} L_2 \non \\
L_3 &=& \frac{1}{4} L_2 \non \\
L_4 &=& -L_3 + R_{abcd} R^{e \hspace{0.1in} c}_{\hspace{0.05in} af}R^{b \hspace{0.1in} e}_{\hspace{0.05in} fd}
\label{addli}
\feqr

\noi one can recover the proposed formula.

\noi \textit{Proof of (12) - (24)} With the assistance of \textit{(7) - (9)} it is a straightforward excercise to show 
their validity.

\subsection{Six dimensional invariants}

In d-dimensions the polynomials  $C^{abcd} C_{cdef} C^{ef}_{\hspace{0.1in} ab}$ and 
$C_{abcd}C^{ebcf} C^{a \hspace{0.1in} d}_{\hspace{0.05in ef}}$ (type-B anomaly according to the geometric 
classification of \cite{deser}) are written as:
\beqr
\textit{$Inv_1$}=C^{abcd} C_{cdef} C^{ef}_{\hspace{0.1in} ab} &=& R^{abcd} R_{cdef} R^{ef}_{\hspace{0.1in} ab} -\frac{12}{d-2}
R_{abcd}R^{abce}R^{d}_{e} + \frac{6}{(d-1)(d-2)} R R^{2}_{abcd} \non \\
&+& \frac{8(2d-3)}{(d-1)^2 (d-2)^3}R^3 - \frac{24(2d-3)}{(d-1)(d-2)^3} R R^{2}_{ab} \non \\
&+& \frac{16(d-1)}{(d-2)^3} R^{ab} R_{bc} R^{c}_{a} 
+ \frac{24}{(d-2)^2} R^{abcd} R_{ac} R_{bd}.
\label{inv1}
\feqr

\beqr
\textit{$Inv_2$}=C_{abcd}C^{ebcf} C^{a \hspace{0.1in} d}_{\hspace{0.05in ef}} &=& \frac{(d^2 +d -4)}{(d-1)^2 (d-2)^3} R^3 - 
\frac{3(d^2 +d -4)}{(d-1)(d-2)^3}RR^{2}_{ab} \non \\
&+& \frac{3}{2(d-1)(d-2)}RR^{2}_{abcd} 
+ \frac{2(3d-4)}{(d-2)^3}R_{ab}R^{bc}R^{a}_{c} \non \\
&+&\frac{3d}{(d-2)^2}R_{abcd}R^{ac}R^{bd} 
- \frac{3}{d-2}R_{abcd}R^{abce}R^{d}_{e} \non \\
&+& R_{abcd} R^{ebcf} R^{a \hspace{0.1in} d}_{\hspace{0.05in} ef}.
\label{inv2}
\feqr

\noi Expressions\rf{inv1} and\rf{inv2} reproduce in $d=6$ the unique already known Weyl invariant polynomials of 
\cite{bonora} constructed only out of Weyl tensors. 
In higher dimensions the independent ways to contract indices among a number of Weyl tensors increases. 

Another invariant made out of covariant derivatives of the Weyl tensor is:
\beqr
\textit{$Inv_3$} &=& -R^3 +8RR^{ab}R_{ab}+2RR^{abcd}R_{abcd}-10R^{ab}R^{c}_{b}R_{ca}-10R_{ab}R^{acdb}R_{cd} \non \\
&+&\frac{1}{2} R\Box R -5 R^{ab}\Box R_{ab}+5 R^{abcd} \Box R_{abcd}.
\label{inv3} 
\feqr

The final nontrivial \footnote{The terminology trivial invariants is adopted to justify the existence of covariant 
total derivatives of polynomials over Riemann tensor and it's covariant derivatives. These invariants coincide 
with variations of all independent local counterterms to effective action.} 
invariant (type-A) we would like to consider stems from the Euler form which exists in any even dimension $d=2n$:
\beqr
E_{2n} &=&\frac{1}{(4\pi)^n \hspace{0.05in} n!} \int_{\mathcal{M}} \epsilon_{a_1 a_2 \cdots a_{2n}} R^{a_1 a_2}\wedge 
\cdots \wedge R^{a_{2n-1}a_{2n}} \non \\
&=& \frac{1}{(4\pi)^n \hspace{0.05in} n!} \frac{1}{2^{n}} \int_{\mathcal{M}} \epsilon_{a_1 a_2 \cdots a_{2n}}
\epsilon^{b_1 b_2 \cdots b_{2n}} R^{a_1 a_2}_{\hspace{0.3in} b_1 b_2} \cdots 
R^{a_{2n-1} a_{2n}}_{\hspace{0.5in} b_{2n-1} b_{2n}} dV \non \\
&=& \frac{1}{(4\pi)^n \hspace{0.05in} n!}\int_{\mathcal{M}} \mathcal{E}_{2n} dV
\label{eulerf}
\feqr 

\noi where 
\beqr
\mathcal{E}_{2n} &=& \frac{1}{2^{n}}\epsilon_{a_1 a_2 \cdots a_{2n}}
\epsilon^{b_1 b_2 \cdots b_{2n}} R^{a_1 a_2}_{\hspace{0.3in} b_1 b_2} \cdots 
R^{a_{2n-1} a_{2n}}_{\hspace{0.5in} b_{2n-1} b_{2n}} \non \\
&=&\frac{1}{2^{n}} \det(\delta^{b_i}_{a_i})
R^{a_1 a_2}_{\hspace{0.3in} b_1 b_2} \cdots R^{a_{2n-1} a_{2n}}_{\hspace{0.5in} b_{2n-1} b_{2n}}; \quad i=1, \cdots , 2n 
\label{e2n}
\feqr
 \noi is the Euler number which is a total divergence, vanishes in all lower integer dimensions and 
$dV=e^1 \wedge \cdots \wedge e^{2n} =\sqrt{-g} d^{2n} x$ is the volume form. In six dimenions the Euler number becomes:
\beqr
\textit{$Inv_4$}=\mathcal{E}_6 &=& R^3 -12 R R^{ab}R_{ab}+16 R^{ab}R^{c}_{b}R_{ac}+24 R^{ab}R^{cd}R_{acbd}
+3RR^{abcd}R_{abcd} \non \\
&+& 4 R^{abcd}R_{ab}^{\hspace{0.1in} ef}R_{cdef}-24R^{ab}R_{a}^{\hspace{0.1in} cde}R_{bcde}
-8R^{abcd}R_{a \hspace{0.1in}c }^{\hspace{0.1in}e \hspace{0.1in} f}R_{bedf}.
\label{e6}
\feqr

\subsection{The vertex $I_2$}

The key idea for calculating the contribution of this vertex is to express each interaction term in terms of polynomials
of the Riemann curvature and its covariant derivatives. This can be achieved by making use of the identities\rf{id1} and
the symmetries implied by the structure of each term separately.

\begin{itemize}

\item With the help of the identity 
$\partial_{a}\bar{\Gamma}^{d}_{b c}=\partial_{a}\bar{\Gamma}^{d}_{(b c)} = -\frac{2}{3} \bar{R}^{d}_{(b c) a}$ one may 
rewrite the second term of $I_2$ as \footnote{Identical results arise if one makes use of the relation
$\partial_{(a}\bar{\Gamma}^{d}_{b ) c}= -\frac{1}{3} \bar{R}^{d}_{(b c a)}$ which communicates with 
the one employed in the text by the relation $\partial_{(a}\bar{\Gamma}^{d}_{b c )}=0$.}:
\beqr
- \frac{1}{3}\bar{R}^{i \hspace{0.15in} j}_{\hspace{0.05in} i_1 i_2}\partial_{i_3}\bar{\Gamma}^{l}_{i k}
\partial_{i_4}\bar{\Gamma}^{k}_{l j} 
= -\frac{4}{27}\bar{R}^{i  \hspace{0.15in} j}_{\hspace{0.05in} i_1 i_2} \bar{R}^{l}_{(i k) i_3}
\bar{R}^{k}_{(l j) i_4}.
\label{term2}
\feqr 

\item Expanding out the identity $\partial_{(a} \partial_{b} \bar{\Gamma}^{e}_{c) d} 
=-\frac{1}{2}D_{(a}\bar{R}^{e}_{b |d| c )}$ we get:
\beqr
\partial_{i_1}\partial_{i_2}\bar{\Gamma}^{l}_{k i} + \partial_{i_1}\partial_{k}\bar{\Gamma}^{l}_{i_2 i}
+ \partial_{i_2}\partial_{k}\bar{\Gamma}^{l}_{i_1 i}
&=& -\frac{1}{4} \big( D_{i_1}\bar{R}^{l}_{i_2 i k} + D_{i_2}\bar{R}^{l}_{i_1 i k} +
D_{k}\bar{R}^{l}_{i_1 i i_2} \non \\
&+& D_{k}\bar{R}^{l}_{i_2 i i_1} + 
D_{i_1}\bar{R}^{l}_{k i i_2} + D_{i_2}\bar{R}^{l}_{k i i_1} \big).
\label{iden2}    
\feqr

\noi  Notice that the third term of $I_2$ is symmetric under the interchange of the lower indices of the Christoffel 
symbols namely  $\mu \leftrightarrow k$ and $l \leftrightarrow \nu$. So in\rf{iden2} we interchange 
these indices and add the two expressions together. We then obtain:
\beqr
&&2 \left( \partial_{i_1}\partial_{i_2}\bar{\Gamma}^{l}_{k i} + \partial_{i_1}\partial_{(k}\bar{\Gamma}^{l}_{i) i_2}
+ \partial_{i_2}\partial_{(k}\bar{\Gamma}^{l}_{i) i_1} \right) \non \\
\!\!\!\!\! &=& \!\!\!\!\!  -\frac{1}{4} \big( D_{i_1}\bar{R}^{l}_{k i i_2} + D_{i_2}\bar{R}^{l}_{k i i_1} +
D_{i_1}\bar{R}^{l}_{i k i_2}+ D_{i_2}\bar{R}^{l}_{i k i_1} \non \\
&+& \!\!\! D_{k}\bar{R}^{l}_{i_2 i i_1} + D_{k}\bar{R}^{l}_{i_1 i i_2} +
 D_{i}\bar{R}^{l}_{i_2 k i_1} + D_{i}\bar{R}^{l}_{i_2 k i_1} \big).
\label{iden3}    
\feqr 

\noi We would like now to express the second and third terms of the lefthand side of\rf{iden3} in terms of 
$\partial_{i_1}\partial_{i_2}\bar{\Gamma}^{l}_{k \mu}$ and covariant derivatives of the Riemann curvature. Starting 
from:
\beqr
D_{i_1} \bar{R}^{l}_{i i_2 k} &=& \partial_{i_1}\partial_{i_2}\bar{\Gamma}^{l}_{i k}
-\partial_{i_1}\partial_{i}\bar{\Gamma}^{l}_{k i_2} \Rightarrow
\non \\
\partial_{i_1}\partial_{(i} \bar{\Gamma}^{l}_{k) i_2}
&=& \partial_{i_1}\partial_{i_2} \bar{\Gamma}^{l}_{(i k)}
-D_{i_1} \bar{R}^{l}_{(i |i_2| k)}
\label{cov1}
\feqr

\noi one has: 
\beqr
\partial_{i_1}\partial_{(k}\bar{\Gamma}^{l}_{i ) i_2}+ \partial_{i_2}\partial_{(k}\bar{\Gamma}^{l}_{i ) i_1}
= 2\partial_{i_1}\partial_{i_2}\bar{\Gamma}^{l}_{k i} - D_{i_1}R^{l}_{(i |i_2| k)} - D_{i_2}R^{l}_{(i |i_1| k)}.
\label{add1}
\feqr

\noi Substituting\rf{add1} into\rf{iden3} one gets:
\beqr
&&  3\partial_{i_1}\partial_{i_2}\bar{\Gamma}^{l}_{k i}  
= -\frac{1}{8} \big( D_{i_1}\bar{R}^{l}_{k i i_2} + D_{i_2}\bar{R}^{l}_{k i i_1} +
D_{i_1}\bar{R}^{l}_{i k i_2}+ D_{i_2}\bar{R}^{l}_{i k i_1} \non \\
&+& \!\!\! D_{k}\bar{R}^{l}_{i_2 i i_1} + D_{k}\bar{R}^{l}_{i_1 i i_2} +
 D_{i}\bar{R}^{l}_{i_2 k i_1} + D_{i}\bar{R}^{l}_{i_2 k i_1} \big) \non \\
&+&  \frac{1}{2}\left(D_{i_1}R^{l}_{i i_2 k} + D_{i_1}R^{l}_{k i_2 i}+ D_{i_2}R^{l}_{i i_1 k} +
D_{i_2}R^{l}_{k i_1 i} \right).
\label{iden4}    
\feqr  

\noi Finally the expression we are going to use for the contribution of the third term in the action $I_2$ is:
\beqr
&&  \partial_{i_1}\partial_{i_2}\bar{\Gamma}^{l}_{k i} 
=-\frac{1}{24} \big( D_{i_1}\bar{R}^{l}_{k i i_2} + D_{i_2}\bar{R}^{l}_{k i i_1} +
D_{i_1}\bar{R}^{l}_{i k i_2}+ D_{i_2}\bar{R}^{l}_{i k i_1} \non \\
&+& \!\!\! D_{k}\bar{R}^{l}_{i_2 i i_1} + D_{k}\bar{R}^{l}_{i_1 i i_2} +
 D_{i}\bar{R}^{l}_{i_2 k i_1} + D_{i}\bar{R}^{l}_{i_2 k i_1} \big) \non \\
&+&  \frac{1}{6}\left(D_{i_1}R^{l}_{i i_2 k} + D_{i_1}R^{l}_{k i_2 i}+ D_{i_2}R^{l}_{i i_1 k} +
D_{i_2}R^{l}_{k i_1 i} \right).
\label{iden5}    
\feqr  

\noi and a similar relation holds for the term $\partial_{i_3}\partial_{i_4}\bar{\Gamma}^{k}_{l i}$. 

\end{itemize} 

\bibliographystyle{plain}

\end{document}

%% file: Definiti.tex
\newcommand{\beq}{\begin{equation}}
\newcommand{\feq}[1]{\label{#1} \end{equation}}
\newcommand{\beqr}{\begin{eqnarray}}
\newcommand{\feqr}{\end{eqnarray}}
\def\non{\nonumber}
\def\noi{\noindent}
\def\slasha#1{\setbox0=\hbox{$#1$}#1\hskip-\wd0\hbox to\wd0{\hss\sl/\/\hss}}

\def\slashb#1{\setbox0=\hbox{$#1$}#1\hskip-\wd0\dimen0=5pt\advance
       \dimen0 by-\ht0\advance\dimen0 by\dp0\lower0.5\dimen0\hbox
         to\wd0{\hss\sl/\/\hss}}
\newcommand{\rf}[1]{(\ref{#1})}
\def\pr{^{\prime}}
\def\np#1#2#3{Nucl. Phys. {\bf{B#1}} (#2) #3}

\def\pr#1#2#3{Phys. Rep. {\bf{#1}} (#2) #3}
\def\prev#1#2#3{Phys. Rev. {\bf{D#1}} (#2) #3}
\def\prl#1#2#3{Phys. Rev. Lett. {\bf{#1}} (#2) #3}
\def\cqg#1#2#3{Class. Quantum Grav. {\bf{#1}} (#2) #3}

\def\ap#1#2#3{Ann. of Phys. {\bf{#1}} (#2) #3}

\def\plb#1#2#3{Phys. Lett. {\bf{B#1}} (#2) #3}

\def\grg#1#2#3{J. Gen. Rel. Grav. {\bf{#1}} (#2) #3}
\def\ijmp#1#2#3{Int. J. Mod. Phys. {\bf{A#1}} (#2) #3}
\def\jdg#1#2#3{J. Diff. Geom. {\bf{#1}} (#2) #3}
\renewcommand{\thefootnote}{\fnsymbol{footnote}}

%% file: Leaf1.tex
\thicklines
\newsavebox{\leaf}
\savebox{\leaf}(10,10){\put(20,0){\oval(20,20)[r]}
\qbezier(0,0)(5,3)(20,10)
\qbezier(0,0)(5,-3)(20,-10)}
\newsavebox{\ninety}
\savebox{\ninety}(0,0){
\rotatebox{90}{\usebox{\leaf}}}
\newsavebox{\onehun}
\savebox{\onehun}(0,0){
\rotatebox{180}{\usebox{\leaf}}}
\newsavebox{\twosev}
\savebox{\twosev}(0,0){
\rotatebox{270}{\usebox{\leaf}}}
\newsavebox{\threehun}
\savebox{\threehun}(0,0){
\rotatebox{360}{\usebox{\leaf}}}
\newsavebox{\fourtyfive}
\savebox{\fourtyfive}(0,0){
\rotatebox{45}{\usebox{\leaf}}}
\newsavebox{\twotwofive}
\savebox{\twotwofive}(0,0){
\rotatebox{225}{\usebox{\leaf}}}
\newsavebox{\thonefive}
\savebox{\thonefive}(0,0){
\rotatebox{315}{\usebox{\leaf}}}